\documentclass[11pt,a4paper]{article}
\usepackage[T1]{fontenc}
\usepackage[utf8]{inputenc}
\usepackage{lmodern}
\usepackage[margin=25mm]{geometry}
\usepackage{amsmath,amssymb,bm}
\usepackage{graphicx}
\usepackage{booktabs,tabularx,array}
\usepackage{microtype}
\usepackage{caption}
\usepackage[section]{placeins}
\usepackage[numbers,sort&compress]{natbib}
\usepackage[hidelinks]{hyperref}
\usepackage{xurl}
\newcommand{\doi}[1]{\href{https://doi.org/#1}{\nolinkurl{#1}}}
\title{Reconstruction of Multiscale Plasma Dynamics Across Operating Regimes}
\author{Maryam Reza\textsuperscript{*}, Farbod Faraji\textsuperscript{**}\\[0.5em]
\small\textsuperscript{*} Independent Researcher\\
\small\textsuperscript{**} Department of Computing, Huxley Building,\\
\small Imperial College London, London SW7 2RH, United Kingdom}
\date{}
\begin{document}
\maketitle

\begin{abstract}
Reconstructing spatially resolved plasma dynamics from a small number of sensors is essential for diagnostics, reduced-order modelling and control, yet remains difficult because the sparse measurements incompletely constrain multiscale, regime-dependent degrees of freedom. The Shallow Recurrent Decoder (SHRED) partially addresses spatial sparsity by using information contained in measurement histories instead; however, its fully connected decoder provides no explicit mechanism for resolving spatial structure across scales or explicit parametric dependency. We introduce the Recurrent Multiscale Affine-modulated Inference Network (ReMAIN), which preserves SHRED's recurrent temporal encoding while replacing its decoder with a U-Net whose feature hierarchy is conditioned by the recurrent state through feature-wise linear modulation. The temporal representation supplies both a dense prior and scale-specific modulation throughout the encoder, bottleneck and decoder. A parametric extension jointly embeds the operating condition and sensor history, enabling reconstruction to adapt as the governing dynamics change with operating regime.

ReMAIN is first benchmarked against the SHRED baseline on six one-dimensional nonlinear Partial Differential Equations (PDEs) representing diverse spatiotemporal dynamics. Across all benchmarks, it reduces reconstruction errors on unseen trajectories and more faithfully resolves sharp transitions, localized extrema and fine-scale variations. The parameter-conditioned model is then demonstrated on a collisionless \(E \times B\ \) plasma subject to perpendicular axial electric and radial magnetic fields, with the electric-field strength serving as the operating parameter. ReMAIN is shown to successfully reconstruct the high-dimensional, multiscale plasma state and recover its regime-dependent spatiotemporal dynamics at electric-field strengths withheld from training. Together, these results show that ReMAIN improves sparse-sensor full-state reconstruction and, through parameter conditioning, generalizes across varying plasma operating regimes.

\end{abstract}

\section{Introduction}

Knowledge of the spatially distributed plasma state is essential for understanding instability and nonlinear plasma dynamics, and for developing predictive models and control strategies \cite{r01}. Yet plasma systems are rarely fully observable. Diagnostics are constrained by accessibility, finite spatial and temporal resolution, cost, and perturbation by intrusive probes \cite{r02}. At the same time, high-fidelity simulations remain computationally demanding because plasma dynamics are nonlinear, multidimensional and multiscale \cite{r01}. These limitations create a persistent gap between the sparse measurements available experimentally and the high-dimensional fields required to characterize the plasma state \cite{r03,r04}.

This problem is particularly relevant in \(E \times B\) plasmas, including Hall thrusters, magnetrons and Penning-type discharges \cite{r05}. Strong gradients and anisotropies excite interacting instabilities and turbulent fluctuations over broad spatial and temporal scales, regulating cross-field electron transport and ultimately device behaviour \cite{r05}. Importantly, these dynamics can vary substantially with operating conditions; changes in electric and magnetic field strengths \cite{r06,r07}, magnetic topology and propellant \cite{r08}, and external forcing \cite{r09} can alter the dominant instabilities, characteristic spectral content, and associated transport. A robust reconstruction method must therefore recover the multiscale plasma structure from sparse measurements across changing physical regimes.

Classical compressed-sensing and state-estimation approaches provide routes to full-state inference, but commonly require sufficiently dense measurements, carefully selected sensor locations, prescribed low-dimensional representations or accurate dynamical models \cite{r10,r11,r12}. These requirements are restrictive in plasma experiments, where diagnostics may be few and their locations dictated by accessibility rather than information-optimal placement. Machine learning (ML) offers an alternative by learning nonlinear mappings between available measurements and the distributed plasma state. However, reconstruction from a few instantaneous measurements remains strongly underdetermined \cite{r10,r11}. Temporal measurement histories provide an additional source of information because signals recorded at accessible locations reflect the same underlying modes, instabilities and transport processes governing the unobserved plasma. Thus, temporal information can partly compensate for missing spatial measurements \cite{r03,r04}, an idea whose theoretical basis has also been examined explicitly in terms of space--time information interchangeability in dynamical systems \cite{r13}. The idea relies on the fact that the transport processes underlying variations in many systems, including plasmas, are intrinsically coupled across space and time -- as reflected in the Partial Differential Equations describing their spatiotemporal evolution. Advection, diffusion, wave propagation, and turbulence are inherently spatiotemporal processes through which spatial structures vary and reconfigure as the system evolves over time. Accordingly, characteristic signatures in temporal measurements can also reveal information about the spatial distributions. A suitably designed ML model can learn to decode these temporal signatures to infer the underlying spatial distributions, including unmeasured quantities and regions in space, to the extent that they are observable from the available measurement histories \cite{r03,r04,r13}.

This principle underlies the Shallow Recurrent Decoder (SHRED) introduced in Ref. \cite{r14}. SHRED combines a recurrent network, typically an LSTM \cite{r15}, that encodes a finite history of sparse measurements with a shallow fully connected decoder that maps the recurrent representation to the high-dimensional state. The original work demonstrated accurate reconstruction from very few sensors with remarkably reduced sensitivity to their placement compared to state-of-the-art approaches \cite{r14}.

SHRED has also been developed specifically for plasma applications. It is demonstrated for reconstruction of complete high-dimensional plasma states from few finite-history measurements of a local plasma quantity, as well as for super-resolution reconstruction of unresolved plasma structure \cite{r03}. SHRED is extended across several \(E \times B\) plasma configurations, reconstructing dynamically coupled plasma fields from minimal local or global measurements and establishing its potential for reduced-order modelling, diagnostics and digital twins \cite{r04,r16,r17}. The methodology has also been applied to reconstruct spatially distributed solar-wind plasma fields from sparse virtual-probe histories \cite{r18} and to infer multiple coupled quantities in a fusion-relevant kinetic plasma from only a few accessible measurements \cite{r02}. These developments establish recurrent sparse sensing as a promising framework for plasma-state inference across a broad range of applications \cite{r03,r04}.

The existing studies nevertheless expose two important limitations of current SHRED application for plasma systems \cite{r03,r04,r18}. First is about the faithful recovery of multiscale spatial structure. Localized, transient, or lower-energy structures can be underrepresented \cite{r18}, particularly when the reconstruction objective is dominated by errors in larger-amplitude features, favouring accurate recovery of the large-scale morphology at the expense of dynamically important smaller-scale fluctuations. In its application to solar-wind plasma reconstruction, for example, the principal large-scale organization was recovered, but sharper transients, higher-frequency variations and small-scale details were more difficult to reconstruct \cite{r18}. This limitation is particularly consequential in plasmas, where instability spectra can span broad ranges of spatial and temporal scales, and comparatively low-energy modes or nonlinear interactions can nevertheless strongly influence the system dynamics \cite{r05,r06,r07}.

Second is regarding the ability to generalize across changing operating regimes \cite{r04}. Previous \(E \times B\) SHRED studies demonstrated good performance on withheld data and different initial conditions \cite{r04}, but also showed that reconstruction becomes more challenging when the underlying dynamics depart from those represented in the training data, motivating explicit parameter conditioning to improve robustness across regimes. This is critical in plasmas because changing an operating parameter may reorganize the dynamics, producing different dominant instabilities, spatial scales and spectra \cite{r06,r07}. Parametric plasma-modelling studies have likewise highlighted the difficulty and importance of learning representations that remain applicable as operating parameters change \cite{r19,r20}.

These limitations point to the spatial decoding stage as a key target for improvement in SHRED. Its recurrent encoder efficiently extracts dynamical information from sparse measurement histories, but the resulting latent state is mapped to the reconstructed field through a shallow fully connected decoder \cite{r03,r14}, which does not explicitly exploit spatial structure across multiple scales.

Here, we introduce the Recurrent Multiscale Affine-modulated Inference Network (ReMAIN). ReMAIN retains the recurrent temporal encoding central to SHRED but replaces the shallow spatial decoder with a multiresolution convolutional U-Net \cite{r21}. The recurrent representation is projected to a dense temporal prior and simultaneously conditions feature processing throughout the U-Net using feature-wise linear modulation (FiLM) \cite{r22}, allowing the temporal context to influence reconstruction at multiple spatial scales. U-Net's skip connections preserve fine- and intermediate-scale information while deeper levels capture progressively larger-scale structure \cite{r21}.

ReMAIN is further extended to parameter-conditioned reconstruction by jointly embedding the operating parameter and measurement history, allowing the reconstructed spatial hierarchy to adapt to the physical regime. The ReMAIN architecture is first compared with a capacity-matched SHRED baseline across six different nonlinear PDE systems; parametric ReMAIN is then applied to a parameter-dependent plasma problem.

\section{Methodology}

\textbf{Architecture overview.} The Recurrent Multiscale Affine-modulated Inference Network (ReMAIN) architecture is designed to recover a high-dimensional state from a finite history of sparse measurements. The model combines three complementary mechanisms: a recurrent encoder that extracts the dynamical information contained in the measurement history, a multiresolution convolutional U-Net that reconstructs the state through hierarchical convolutional features, and feature-wise linear modulation (FiLM) that conditions the reconstruction process on the recurrent temporal representation.

While temporal information extraction and full-state reconstruction remain distinct, they are closely coupled. The finite measurement history is first processed by the recurrent encoder to obtain a compact latent representation of its temporal evolution. This representation is linearly projected to form a dense temporal prior, which is combined with the measurements at the final time point of the input history, their sampling mask, and fixed positional information to form the input to the U-Net reconstruction decoder. In parallel, the same temporal representation is used by the FiLM layers to condition the feature maps throughout the U-Net encoder, bottleneck, and decoder, so that feature processing at different levels of the reconstruction network can adapt to the temporal context inferred from the measurement history.

\textbf{Recurrent temporal representation.} Let \(\mathbf{y}_{t} \in R^{m}\ \) denote the measurements from \(m\) sensors at time \(t\), and consider a time history of measurements of length \(L\),

\begin{equation}
\label{eq:1}
\mathbf{Y}_{t} = \left\lbrack \mathbf{y}_{t - L + 1},\mathbf{y}_{t - L + 2},\ldots,\mathbf{y}_{t} \right\rbrack \in \mathbb{R}^{L \times m}.
\end{equation}

The complete measurement history is processed by a stacked long short-term memory (LSTM) network \cite{r15}. Denoting the recurrent mapping by \(\mathcal{R}_{\theta}\),

\begin{equation}
\label{eq:2}
\mathbf{z}_{t} = \mathcal{R}_{\theta}\left( \mathbf{Y}_{t} \right),\ \ \mathbf{z}_{t} \in \mathbb{R}^{h},
\end{equation}

where \(h\) is the LSTM hidden dimension. In the current implementation, two LSTM layers are considered. The representation \(\mathbf{z}_{t}\) is taken from the final hidden state of the last recurrent layer. The vector therefore provides a compact representation of the recent temporal evolution observed by all sensors. It acts as a global dynamical context from which the reconstruction decoder network is conditioned.

\textbf{Dense temporal prior and measurement embedding.} The temporal representation is first mapped through a learned linear transformation to the \(N\)-dimensional state space

\begin{equation}
\label{eq:3}
\mathbf{p}_{t} = \mathbf{W}_{p}\mathbf{z}_{t} + \mathbf{b}_{p},\ \ \mathbf{p}_{t} \in \mathbb{R}^{N}.
\end{equation}

This quantity can be interpreted as a dense temporal prior: a preliminary state-level representation inferred solely from the recent measurement history.

The current measurements are supplied independently to the reconstruction network. Let \(\mathbf{P} \in \left\{ 0,1 \right\}^{m \times N}\) be the sampling operator that extracts the observed components from a complete state. A sparse measurement canvas can then be written as

\begin{equation}
\label{eq:4}
\mathbf{s}_{t} = \mathbf{P}^{T}\mathbf{y}_{t},
\end{equation}

while the associated sampling mask is

\begin{equation}
\label{eq:5}
\mathbf{q} = \mathbf{P}^{T}\mathbf{1}_{m}.
\end{equation}

Thus, \(\mathbf{s}_{t}\) identifies the measured values at their corresponding reconstruction locations, whereas \(\mathbf{q}\) distinguishes measured locations from unobserved locations. This explicit distinction is important because an unobserved entry must not be interpreted by the network as a physical measurement equal to zero.

The input to the convolutional reconstruction network is formed by concatenating the dense temporal prior \(\mathbf{p}_{t}\), the current sparse observations \(\mathbf{s}_{t}\), the sensor mask \(\mathbf{q}\), and two fixed coordinate-based positional-encoding channels \(\mathbf{c}\), defined over all reconstruction locations,

\begin{equation}
\label{eq:6}
\mathbf{X}_{t}^{(0)} = \operatorname{Concat}\left( \mathbf{p}_{t},\mathbf{s}_{t},\mathbf{q},\mathbf{c} \right).
\end{equation}

The positional encoding channels provide the convolutional decoder with explicit information about the location of each reconstruction point. They are distinct from the sampling mask \(\mathbf{q}\), which identifies only the locations at which measurements are available.

\textbf{Multiresolution convolutional U-Net representation.} The reconstruction component follows a multiresolution encoder-bottleneck-decoder structure \cite{r21} comprising three encoder levels and a bottleneck. If the number of feature channels at the first encoder level is denoted by \(C\) , the channel dimensions increase by a factor of two at each successive level, giving \(C\), \(2C\), \(4C\) across the encoder, followed by \(8C\) at the bottleneck.

Each convolutional block \(C_{l}^{E},\ \ l = 1,2,3\) consists of two local convolutions. Each convolution is followed by group normalization and a GELU nonlinear activation, with dropout included between the two convolutional operations. The implementation uses kernel width three for the local convolutional operations.

The encoder progressively transforms the input while reducing its resolution. Let \(D\) denote downsampling, and \(\mathcal{F}_{l}^{E}\) denote the temporal FiLM conditioning described in the following subsection. In the present architecture, \(D\ \) is implemented by factor-of-two max pooling, progressively reducing the feature resolution between successive encoder levels. The encoder is written as

\begin{equation}
\label{eq:7}
\mathbf{e}_{1} = \mathcal{F}_{ 1}^{E}\left( C_{1}^{E}\left( \mathbf{X}_{t}^{(0)} \right),\mathbf{z}_{t} \right),
\end{equation}

\begin{equation}
\label{eq:8}
\mathbf{e}_{l} = \mathcal{F}_{l}^{E}\left( C_{l}^{E}\left( D\left( \mathbf{e}_{l - 1} \right) \right),\mathbf{z}_{t} \right),\ \ l = 2,3,
\end{equation}

\begin{equation}
\label{eq:9}
\mathbf{b} = \mathcal{F}^{B}(C^{B}(D(\mathbf{e}_{3})),\mathbf{z}_{t}),
\end{equation}

Here, \(\mathbf{e}_{1},\ \mathbf{e}_{2},\ \) and \(\ \mathbf{e}_{3}\) are the feature maps generated at the three encoder resolutions, and \(\mathbf{b}\) denotes the lowest-resolution bottleneck representation. Each downsampling step reduces the feature resolution while increasing the number of channels, thereby enlarging the effective receptive field of the deeper representations and allowing increasingly coarse-scale information to be captured. The encoder feature maps are retained and transferred directly to the corresponding decoder levels through U-Net skip connections. These connections prevent all information from being forced through the lowest-resolution bottleneck. Instead, features obtained at finer and intermediate resolutions remain directly available during reconstruction and are combined with the progressively upsampled decoder representation.

The decoder reverses this multiresolution process. At each decoder level, the representation arriving from the preceding coarser level is first upsampled by interpolation to match the resolution of the corresponding encoder feature map. A pointwise convolution is then applied to adjust the number of feature channels before concatenation with the skip-connected encoder representation at the same resolution. The resulting combined representation is processed by a convolutional block and subsequently conditioned by FiLM.

Let \(U\) denote the upsampling operator, \(Q_{l}\) the pointwise channel projection, and \(\lbrack.,.\rbrack\ \) channel-wise concatenation. The decoder stages are therefore

\begin{equation}
\label{eq:10}
\mathbf{d}_{3} = \mathcal{F}_{3}^{D}\left( C_{3}^{D}\left( \left\lbrack Q_{3}\left( U\left( \mathbf{b} \right) \right),\mathbf{e}_{3} \right\rbrack \right),\mathbf{z}_{t} \right).
\end{equation}

\begin{equation}
\label{eq:11}
\mathbf{d}_{l} = \mathcal{F}_{l}^{D}\left( C_{l}^{D}\left( \left\lbrack Q_{l}\left( U\left( d_{l + 1} \right) \right),\mathbf{e}_{l} \right\rbrack \right),\mathbf{z}_{t} \right),\ \ l = 2,1.
\end{equation}

Each decoder level therefore combines information from two complementary pathways: the progressively reconstructed representation arriving from the coarser level and the encoder features retained at the same resolution. The skip connections thereby retain finer-resolution features that would otherwise need to be recovered solely from the bottleneck representation.

Following the final decoder level, the highest-resolution feature representation \(\mathbf{d}_{1}\) is mapped to the reconstructed state by a pointwise output projection,

\begin{equation}
\label{eq:12}
{\widehat{\mathbf{x}}}_{t} = \mathbf{W}_{o}\left( \mathbf{d}_{1} \right),\ \ {\widehat{\mathbf{x}}}_{t} \in \mathbb{R}^{N}
\end{equation}

The output projection reduces the decoder feature channels to the required reconstructed-state output.

\textbf{FiLM conditioning by the recurrent state.} The recurrent temporal representation is used not only to construct the dense temporal prior, but also to condition the feature processing throughout the U-Net. This conditioning is performed using feature-wise linear modulation (FiLM) \cite{r22}, which applies a learned channel-wise affine transformation to the feature maps at each encoder level, the bottleneck, and each decoder level.

For either an encoder or decoder level, let \(S \in \{ E,D\}\) identify the corresponding branch and let \(\mathbf{H}_{l}^{S}\) denote the feature map produced by the convolutional block at level \(l\), with \(C_{l}^{S}\) feature channels. The recurrent representation \(\mathbf{z}_{t}\) is mapped through a learned affine transformation to two channel-wise modulation vectors,

\begin{equation}
\label{eq:13}
\left\lbrack \begin{array}{r}
\mathbf{\gamma}_{l}^{S} \\
\mathbf{\beta}_{l}^{S}
\end{array} \right\rbrack = \mathbf{W}_{F,l\ }^{S}\mathbf{z}_{t} + \mathbf{b}_{F,l}^{S},
\end{equation}

where \(\mathbf{\gamma}_{l}^{S},\ \mathbf{\beta}_{l}^{S} \in \mathbb{R}^{C_{l}^{S}}\). The FiLM transformation is then defined as

\begin{equation}
\label{eq:14}
\mathcal{F}_{l}^{S}\left( \mathbf{H}_{l}^{S}\mathbf{,}\mathbf{z}_{t}\mathbf{} \right)\mathbf{=}\mathbf{H}_{l}^{S}\mathbf{\odot}\left( 1 + \mathbf{\gamma}_{l}^{S}\mathbf{} \right)\mathbf{+}\mathbf{\beta}_{l}^{S},
\end{equation}

where \(\mathbf{\odot}\) denotes channel-wise multiplication. The vectors \(\mathbf{\gamma}_{l}^{S}\) and \(\ \mathbf{\beta}_{l}^{S}\) are broadcast over all locations of the corresponding feature map. Thus, each feature channel is independently scaled and shifted according to the temporal representation \(\mathbf{z}_{t}\). The factor \(1 + \mathbf{\gamma}_{l}^{S}\) gives the scaling transformation an identity-centered form.

An analogous transformation is applied to the bottleneck,

\begin{equation}
\label{eq:15}
\left\lbrack \begin{array}{r}
\mathbf{\gamma}^{B} \\
\mathbf{\beta}^{B}
\end{array} \right\rbrack = \mathbf{W}_{F\ }^{B}\mathbf{z}_{t} + \mathbf{b}_{F}^{B},
\end{equation}

and

\begin{equation}
\label{eq:16}
\mathcal{F}^{B}\left( \mathbf{H}^{B}\mathbf{,}\mathbf{z}_{t}\mathbf{} \right)\mathbf{=}\mathbf{H}^{B}\mathbf{\odot}\left( 1 + \mathbf{\gamma}^{B}\mathbf{} \right)\mathbf{+}\mathbf{\beta}^{B}.
\end{equation}

\begin{figure}[htbp]
\centering
\includegraphics[width=.78\linewidth,height=0.78\textheight,keepaspectratio]{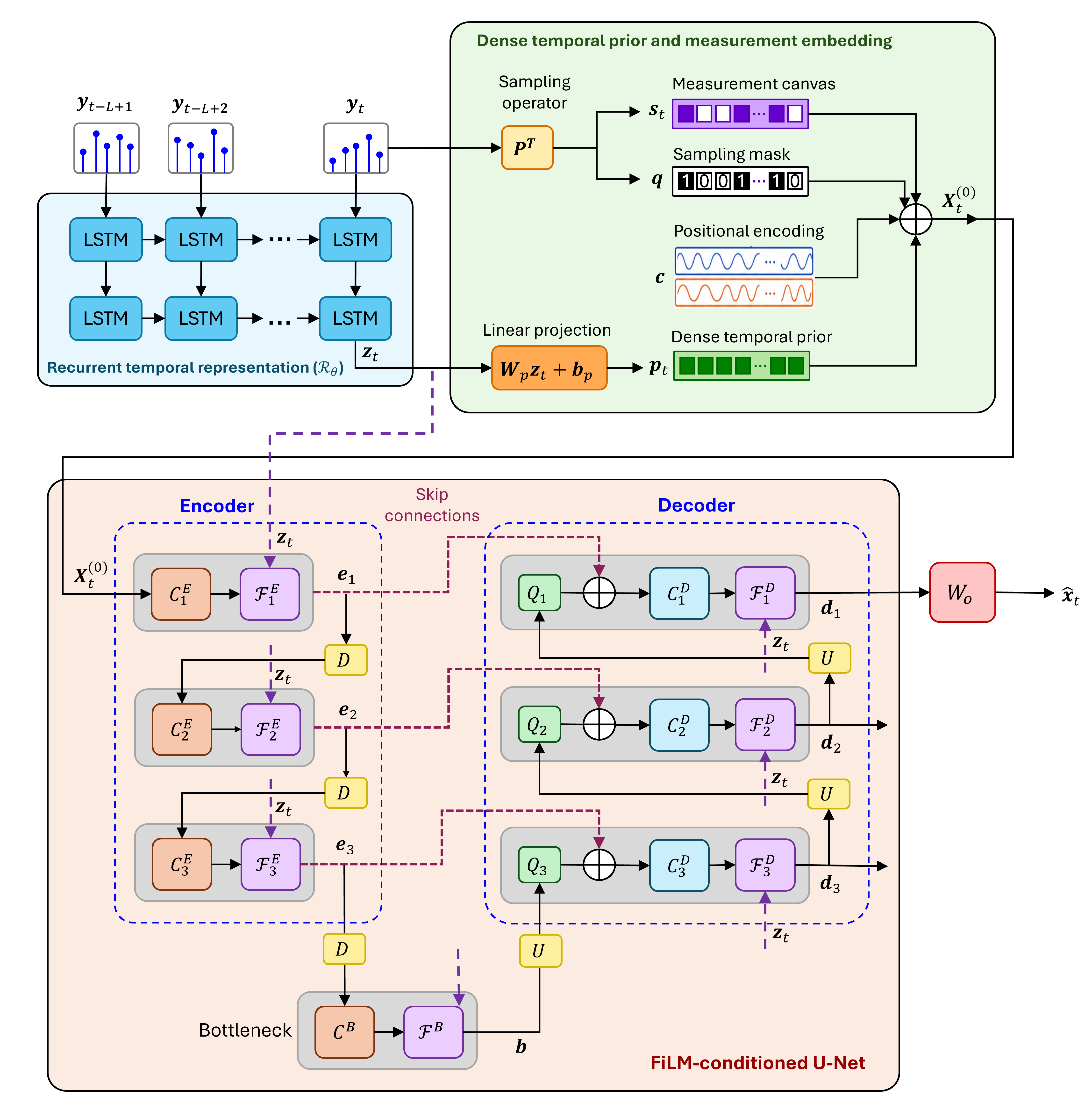}
\caption{Schematic of the Recurrent Multiscale Affine-modulated Inference Network (ReMAIN) architecture.}
\label{fig:1}
\end{figure}

Accordingly, the same recurrent representation \(\mathbf{z}_{t}\) conditions the reconstruction at every resolution of the U-Net, but each level possesses its own learned FiLM transformation. The recurrent state therefore provides a common temporal context, while the independently learned scale and shift parameters allow that context to influence different feature channels and resolution levels differently. FiLM is applied after the convolutional processing at all three encoder levels and the bottleneck, and after convolutional processing of the skip-connected features at all three decoder levels. The ReMAIN architecture is schematically illustrated in Figure~\ref{fig:1}.

\textbf{Relation to baseline Shallow Recurrent Decoder (SHRED).} The baseline SHRED architecture is used for comparison of the performance of ReMAIN. In SHRED, the measurement history is processed by a recurrent network, and the final recurrent hidden state is passed directly through a shallow fully connected decoder \cite{r14}. The implemented baseline consists specifically of the two LSTM layers followed by two hidden fully connected layers and a final output layer,

\begin{equation}
\label{eq:17}
\mathbf{z}_{t} = \mathcal{R}_{\theta}\left( \mathbf{Y}_{t} \right)
\end{equation}

\begin{equation}
\label{eq:18}
{\widehat{\mathbf{x}}}_{t} = \mathbf{W}_{3}\sigma\lbrack\mathbf{W}_{2}\sigma(\mathbf{W}_{1}\mathbf{z}_{t} + \mathbf{b}_{1}) + \mathbf{b}_{2}\rbrack + \mathbf{b}_{3},
\end{equation}

where \(\sigma\) denotes the decoder ReLU activation function.

The ReMAIN architecture retains the central SHRED principle that a finite measurement history can be compressed into a recurrent representation suitable for full-state reconstruction. The principal modification concerns how the recurrent representation is converted into the reconstructed state.

\begin{figure}[htbp]
\centering
\includegraphics[width=1\linewidth,height=0.78\textheight,keepaspectratio]{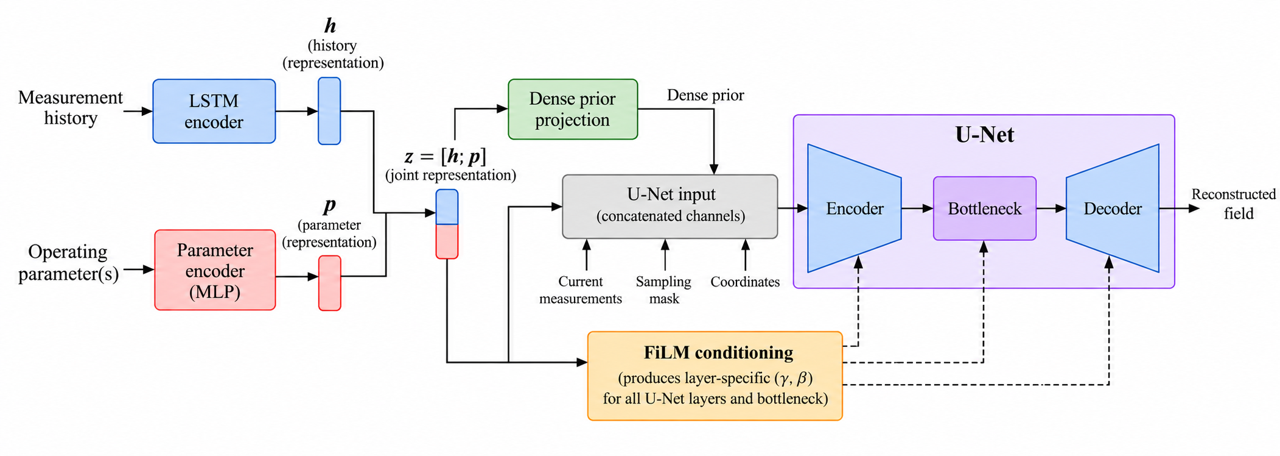}
\caption{Schematic architecture of the parameter-conditioned ReMAIN.}
\label{fig:2}
\end{figure}

\textbf{Parametric ReMAIN.} ReMAIN is extended to incorporate operating parameter(s) as additional conditioning variable(s). The parameter is encoded separately from the measurement history through a learned fully connected embedding network, and the resulting parameter representation is concatenated with the final LSTM hidden state to form a joint history--parameter latent representation. This joint representation replaces the history-only recurrent state as the conditioning variable for both the learned dense prior and the FiLM transformations throughout the reconstruction network. The resulting architecture therefore conditions the reconstruction simultaneously on the observed temporal evolution and the prescribed operating parameter, while retaining the recurrent encoding, dense-prior construction, multiscale reconstruction, and FiLM-conditioning mechanisms described in Section 2. The parametric extension of ReMAIN is schematically illustrated in Figure~\ref{fig:2}.

\section{Comparison of ReMAIN and SHRED using numerical PDE benchmark problems}

Six one-dimensional nonlinear PDE systems were used to evaluate sparse-sensor state reconstruction: KdV--Kuramoto--Sivashinsky (KdV--KS), stochastically forced viscous Burgers, isothermal compressible Navier--Stokes, FitzHugh--Nagumo, driven damped sine--Gordon and complex Ginzburg--Landau (CGL). All systems were discretized on a uniform grid with \(N_{x} = 128\) and periodic boundary conditions. The PDE equations for these cases are as follows.

\textbf{Korteweg-de Vries--Kuramoto-Sivashinsky (KdV-KS) equation:}

\begin{equation}
\label{eq:19}
u_{t} + \ c_{a}dv\ uu_{x} + \beta u_{xxx} + u_{xx} + \nu_{4}u_{xxxx} = 0,
\end{equation}

with \(c_{a}dv = 1,\ \ \beta = 0.1,\ \ \nu_{4} = 1\), and \(x \in \lbrack 0,60)\).

\textbf{Forced Burgers equation:}

\begin{equation}
\label{eq:20}
u_{t} + c_{a}dv\ uu_{x} = \nu u_{xx} + f(x,t),
\end{equation}

with \(c_{a}\mathrm{dv} = 1\), \(\nu = 0.05\), \(x \in \lbrack 0,2\pi)\).

\textbf{Isothermal compressible Navier-Stokes system:}

\begin{equation}
\label{eq:21}
\rho_{t} + (\rho u)_{x} = 0,
\end{equation}

\begin{equation}
\label{eq:22}
(\rho u)_{t} + (\rho u^{2} + c_{s}^{2}\rho)_{x} = (\rho\nu u_{x})_{x} + \rho f(x,t),
\end{equation}

with \(c_{s} = 1,\ \ \nu = 0.01\), and \(x \in \lbrack 0,4\pi)\). The reconstructed quantity was velocity \(\ u\).

\textbf{FitzHugh-Nagumo system:}

\begin{equation}
\label{eq:23}
u_{t} = \ D_{u}u_{xx} + u - u^{3}/3 - v + I(x,t),
\end{equation}

\begin{equation}
\label{eq:24}
v_{t} = D_{v}v_{xx} + \varepsilon(u + a - bv),
\end{equation}

with \(D_{u} = 0.10,\ \ D_{v} = 0.01,\ \ \varepsilon = 0.08,\ \ a = 0.70,\ \ b = 0.80,\) and \(x \in \lbrack 0,50)\). The activator field \(u\) was reconstructed.

\textbf{Driven damped sine-Gordon system:}

\begin{equation}
\label{eq:25}
u_{t} = v,
\end{equation}

\begin{equation}
\label{eq:26}
v_{t} = u_{xx} - \alpha v - \lbrack 1 + A_{p}\ \cos(\omega_{p}t)\rbrack\ \sin\ u\  + A_{d}\ \cos(2\pi x/L - \omega_{d}t + \varphi_{d}),
\end{equation}

with \(L = 32,\ \ \alpha = 0.10,\ \ A_{p} = 0.30,\ \ \omega_{p} = 1.50,\ \ A_{d} = 0.40\), and \(\omega_{d} = 0.80\), and \(x\  \in \ \lbrack 0,32)\). The displacement \(u\) was reconstructed.

\textbf{Complex Ginzburg--Landau (CGL) equation}:

\begin{equation}
\label{eq:27}
A_{t} = A + (1 + ic_{1})A_{xx} - (1 - ic_{3})|A|^{2}A,
\end{equation}

with \(c_{1} = c_{3} = 2\), and \(x \in \lbrack 0,100)\). The reconstructed variable was \(\operatorname{Re}(A)\).

The PDEs were advanced using problem-appropriate spectral or finite-volume discretizations and time integrators, with internal solver time steps finer than the retained snapshot intervals. The numerical schemes, solver time steps, and temporal sampling intervals used for each benchmark are summarized in Table~\ref{tab:schemes}.

\begin{table}[htbp]
\centering
\small
\setlength{\tabcolsep}{4pt}
\renewcommand{\arraystretch}{1.22}
\begin{tabularx}{\linewidth}{@{}>{\raggedright\arraybackslash}p{0.175\linewidth}>{\raggedright\arraybackslash}X>{\raggedright\arraybackslash}p{0.125\linewidth}>{\centering\arraybackslash}p{0.075\linewidth}>{\centering\arraybackslash}p{0.12\linewidth}>{\centering\arraybackslash}p{0.12\linewidth}@{}}
\toprule
\textbf{PDE} & \textbf{Spatial discretization} & \textbf{Time integrator} & \textbf{Solver $\Delta t$} & \textbf{Sampling stride [time steps]} & \textbf{Sampling interval [time unit]} \\
\midrule
CGL & Fourier spectral & ETDRK4 \cite{cox2002} & 0.01 & 10 & 0.10 \\
KdV--KS & Fourier pseudospectral; 2/3 de-aliasing & ETDRK4 \cite{cox2002} & 0.05 & 5 & 0.25 \\
Forced Burgers & Fourier pseudospectral; 2/3 de-aliasing & ETDRK4 \cite{cox2002} & 0.005 & 20 & 0.10 \\
Compressible Navier--Stokes & Finite volume; Rusanov flux & SSP-RK3 & 0.001 & 30 & 0.03 \\
FitzHugh--Nagumo & Fourier spectral Laplacian & Classical RK4 & 0.02 & 10 & 0.20 \\
Driven sine--Gordon & Fourier spectral Laplacian & Classical RK4 & 0.01 & 10 & 0.10 \\
\bottomrule
\end{tabularx}
\caption{Spatial discretization, time-integration scheme, solver time step, and temporal sampling parameters used for the six PDE benchmarks.}
\label{tab:schemes}
\end{table}

Independent trajectories were generated by resampling the initial conditions and, where applicable, the forcing or drive parameters. For KdV--KS, forced Burgers, FitzHugh--Nagumo and sine--Gordon, smooth initial fields were constructed from the first eight Fourier modes \(m = 1,2\ldots 8\) with Gaussian modal amplitudes, \(a_{m}\sim N\ (0,1/m)\), and phases sampled from a uniform distribution \(\phi_{m}\sim U(0,\ 2\pi)\), followed by case-specific normalization. KdV--KS and forced Burgers were rescaled to zero mean and standard deviation \(0.5\); the FitzHugh--Nagumo activator perturbation had amplitude \(0.15\) about \(u = - 1\); and sine--Gordon displacement and velocity were rescaled to standard deviations 0.20 and 0.10.

Compressible Navier--Stokes used independently generated low-wavenumber Fourier perturbations for density and velocity, rescaled to density standard deviation 0.10 about mean 1 and velocity standard deviation 0.20. For CGL, the initial complex field amplitude was sampled over \(\lbrack 0.7,1.1\rbrack\) and the phase over \(\lbrack - \pi,\ \pi)\).

For forced Burgers and compressible Navier--Stokes, independent stochastic forcing was generated for every trajectory using low-order Fourier modes with Ornstein--Uhlenbeck coefficients. Forced Burgers used modes \(m = 1,2,3\), correlation time \(0.5\), and forcing RMS amplitude \(0.25\); compressible Navier--Stokes used modes \(m = 1\), \(2\), correlation time \(4\), and forcing RMS amplitude \(4\).

For each PDE, 300 independent trajectories were generated. Each trajectory was first evolved through a 10,000-step burn-in using the corresponding solver time step, after which 1000 post-transient snapshots were retained. 70\% of the trajectories were used for training and the remaining 30\% were held out for testing. The split was therefore performed at the trajectory level, so that no trajectory contributed to both sets. Test trajectories corresponded to unseen initial conditions; for the forced systems they additionally contained independent forcing or drive realizations.

Sparse measurements were taken at fixed equidistant spatial locations spanning the computational grid, including the first and last grid points. \(10\) sensors were used for all PDEs except compressible Navier--Stokes, for which \(15\) sensors were used. Each input sample consisted of a finite history of sensor measurements, while the target was the complete 128-point state at the final time of that history.

The temporal history length was selected independently for each PDE from the autocorrelation structure of the training data. Normalized autocorrelation functions were computed for each spatial location and averaged across all training trajectories and locations. Two characteristic timescales were then extracted: the first \(\tau_{1/e}\) at which the mean autocorrelation magnitude fell below \(1/e\), and the integrated autocorrelation time \(\tau_{int}\), which measures the cumulative temporal persistence of correlations and was estimated using the initial-positive-pair method. The history span was set by the larger of these two timescales, with a 20\% margin

\begin{equation}
\label{eq:28}
L_{h}\  = \ \lceil 1.2\ \max(\tau_{1/e},\ \ \tau_{int})\rceil\  + \ 1.
\end{equation}

This produced history lengths of 53, 77, 109, 52, 102 and 28 snapshots for KdV-KS, forced Burgers, compressible Navier-Stokes, FitzHugh-Nagumo, sine-Gordon, and CGL respectively.

All fields were scaled using a feature-wise min-max transformation fitted exclusively to the training trajectories and subsequently applied unchanged to the test data.

\subsection{Model settings and training procedure}

Baseline SHRED and ReMAIN were trained independently using identical sensor histories, reconstruction targets, and train--test data splits. Both models used a two-layer LSTM with 64 hidden units. In baseline SHRED, the final recurrent state was passed through two fully connected hidden layers of width 725 and 800, respectively, each using ReLU activation, followed by the full-state output layer. ReMAIN used the same recurrent encoder but replaced the fully connected decoder with a FiLM-conditioned one-dimensional U-Net. The recurrent state was projected to a dense state-level temporal prior and simultaneously used to modulate the U-Net feature maps through independent FiLM transformations at each encoder level, bottleneck, and decoder level, as described in Section 2. The U-Net comprised three encoder levels and a bottleneck, with each convolutional block containing two successive one-dimensional convolutions of kernel size 3, each followed by Group Normalization and GELU activation, with dropout applied within the block. Circular padding was used for the convolutional operations to respect the periodic spatial domains, and sine/cosine coordinate channels provided positional information. Factor-of-two max pooling was used between encoder levels, while skip connections and interpolation-based upsampling were used in the decoder. The channel width doubled between successive levels.

A base width of 32, giving 32--64--128--256 channels, was used for CGL, KdV--KS, Forced Burgers, compressible Navier--Stokes, FitzHugh--Nagumo and sine--Gordon. These settings yielded closely matched model capacities: 783,189 versus 782,817 trainable parameters for SHRED and ReMAIN, respectively, for CGL, KdV--KS, Forced Burgers, FitzHugh--Nagumo and sine--Gordon, and 784,469 versus 784,097 for compressible Navier--Stokes. The slightly larger counts for compressible Navier--Stokes arise solely from the larger number of input sensors. This parameter matching ensures a fair comparison, so that differences in performance can be attributed primarily to the decoder architecture rather than differences in model capacity.

Sensor values were not imposed explicitly on the reconstructed output; thus both models learned the complete field, including values at observed locations.

Both models were optimized using Adam with an initial learning rate of \(10^{-3}\), weight decay \(10^{-5}\), Smooth-\(L_{1}\) loss with \(\beta = 0.02\), and gradient-norm clipping at \(1.0\). The learning rate was controlled by ReduceLROnPlateau, using a reduction factor of \(0.5\), patience of five monitoring evaluations, and a \(5\%\) relative-improvement threshold. Training was performed for \(300\) epochs, with reconstruction performance evaluated every 10 epochs. Batch size was 512 for all cases.

Predictions on the held-out test trajectories were inverse-transformed to physical units before evaluation. For each test trajectory, reconstruction accuracy was quantified by the trajectory-level relative \(L_{2}\) error,

\begin{equation}
\label{eq:29}
\varepsilon\  = \frac{\left\| {\widehat{U}}_{r}\  - \ U_{r} \right\|_{2}}{\left\| U_{r} \right\|_{2}}
\end{equation}

where \({\widehat{U}}_{r}\) and \(U_{r}\) are predicted and ground truth snapshots, respectively. This produced 90 independent test-error values for each PDE and model. Because the test set consisted of complete trajectories excluded from training, the reported errors quantify generalization to unseen dynamical realizations.

\subsection{Results}

\begin{figure}[htbp]
\centering
\includegraphics[width=1\linewidth,height=0.78\textheight,keepaspectratio]{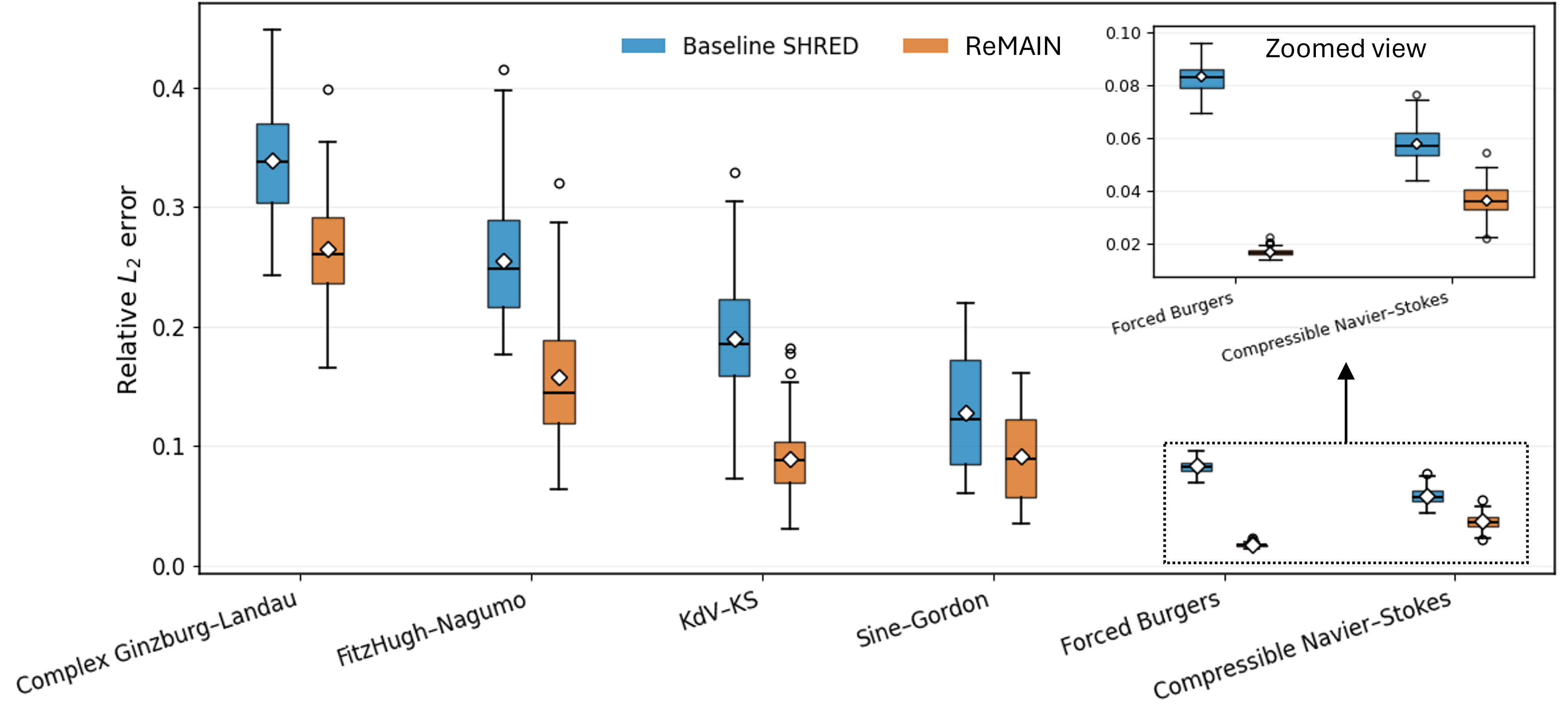}
\caption{Test relative-\(L_{2}\) error distributions for SHRED and ReMAIN across the PDE benchmarks. Boxes show the interquartile range (25th--75th percentiles) of the relative \(L_{2}\) error across test snapshots; the black horizontal line denotes the median and the white diamond is the mean. The vertical lines extend to the most extreme non-outlier values lying no more than 1.5 times the box height below the lower quartile or above the upper quartile, and white circles indicate outliers beyond this range.}
\label{fig:3}
\end{figure}

Across all PDE benchmarks, ReMAIN consistently reduced the reconstruction error relative to baseline SHRED, with the improvement extending across the distributions of held-out trajectories as seen in Figure~\ref{fig:3}. Because the two models use the same recurrent encoding principle and closely matched parameter counts, this systematic improvement indicates that the multiscale, FiLM-conditioned reconstruction stage provides additional capability beyond the shallow fully connected decoder.

\begin{figure}[htbp]
\centering
\includegraphics[width=1\linewidth,height=0.78\textheight,keepaspectratio]{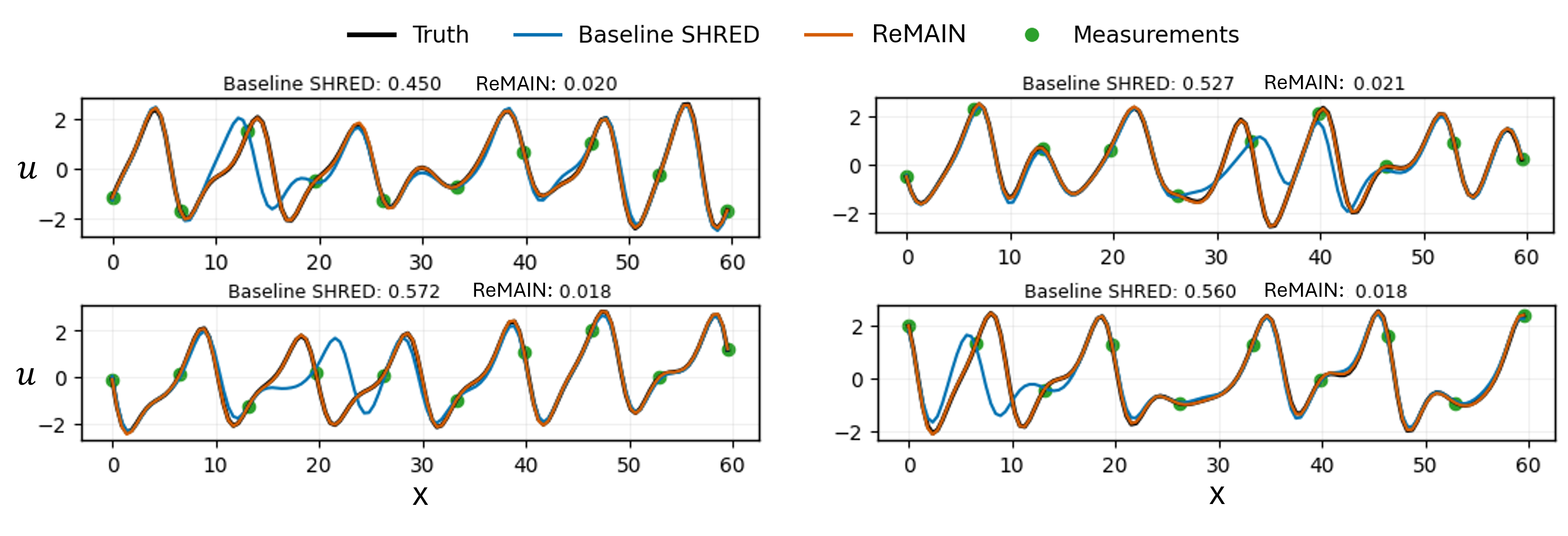}
\caption{Comparison of SHRED and ReMAIN reconstructions for representative held-out test snapshots of the KdV--KS system. The green markers indicate sensor locations. Values above each subplot indicate the relative \(L_{2}\) reconstruction errors of SHRED and ReMAIN for the displayed snapshot.}
\label{fig:4}
\end{figure}

\begin{figure}[htbp]
\centering
\includegraphics[width=1\linewidth,height=0.78\textheight,keepaspectratio]{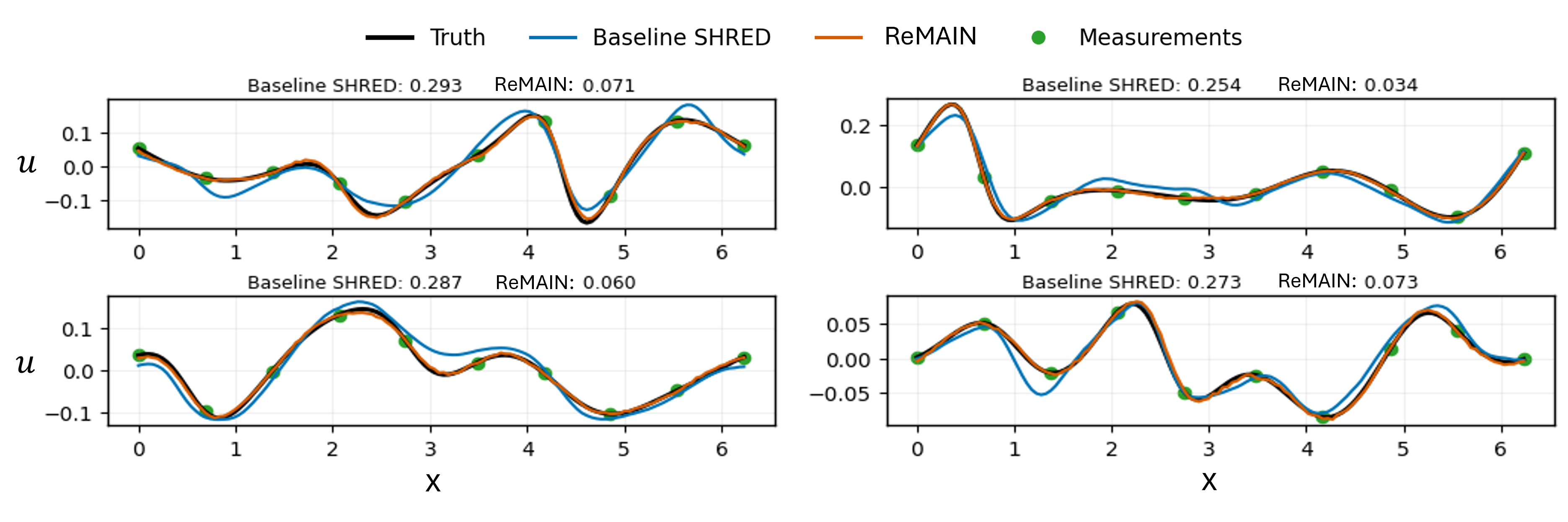}
\caption{Comparison of SHRED and ReMAIN reconstructions for representative held-out test snapshots of the driven Forced Burgers system. Values above each subplot indicate the relative \(L_{2}\) reconstruction errors of SHRED and ReMAIN for the displayed snapshot.}
\label{fig:5}
\end{figure}

\begin{figure}[htbp]
\centering
\includegraphics[width=1\linewidth,height=0.78\textheight,keepaspectratio]{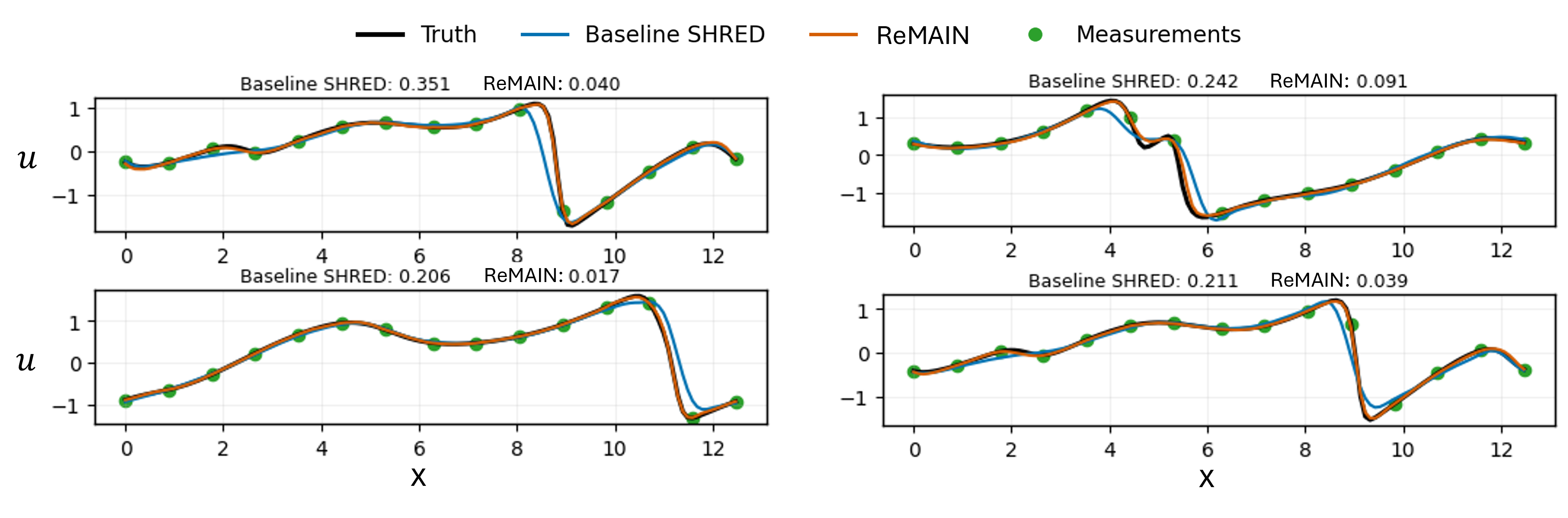}
\caption{Comparison of SHRED and ReMAIN reconstructions for representative held-out test snapshots of the driven Compressible Navier-Stokes system. Values above each subplot indicate the relative \(L_{2}\) reconstruction errors of SHRED and ReMAIN for the displayed snapshot.}
\label{fig:6}
\end{figure}

\begin{figure}[htbp]
\centering
\includegraphics[width=1\linewidth,height=0.78\textheight,keepaspectratio]{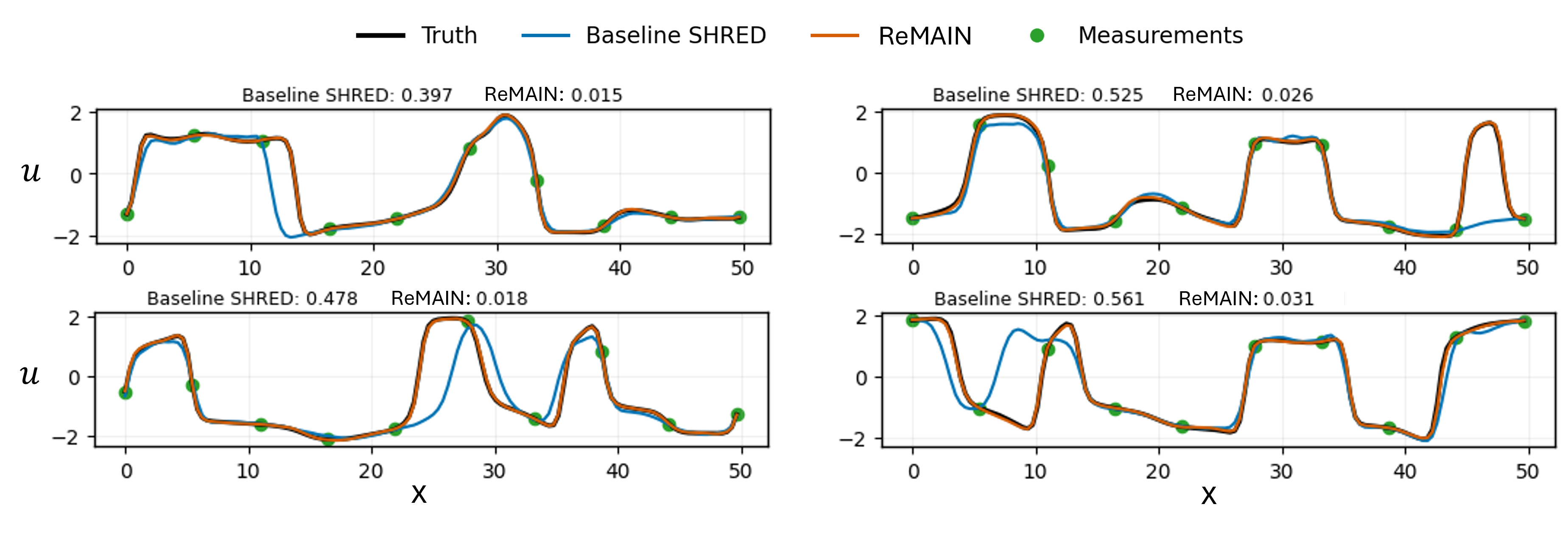}
\caption{Comparison of SHRED and ReMAIN reconstructions for representative held-out test snapshots of the driven FitzHugh-Nagumo system. Values above each subplot indicate the relative \(L_{2}\) reconstruction errors of SHRED and ReMAIN for the displayed snapshot.}
\label{fig:7}
\end{figure}

\begin{figure}[htbp]
\centering
\includegraphics[width=1\linewidth,height=0.78\textheight,keepaspectratio]{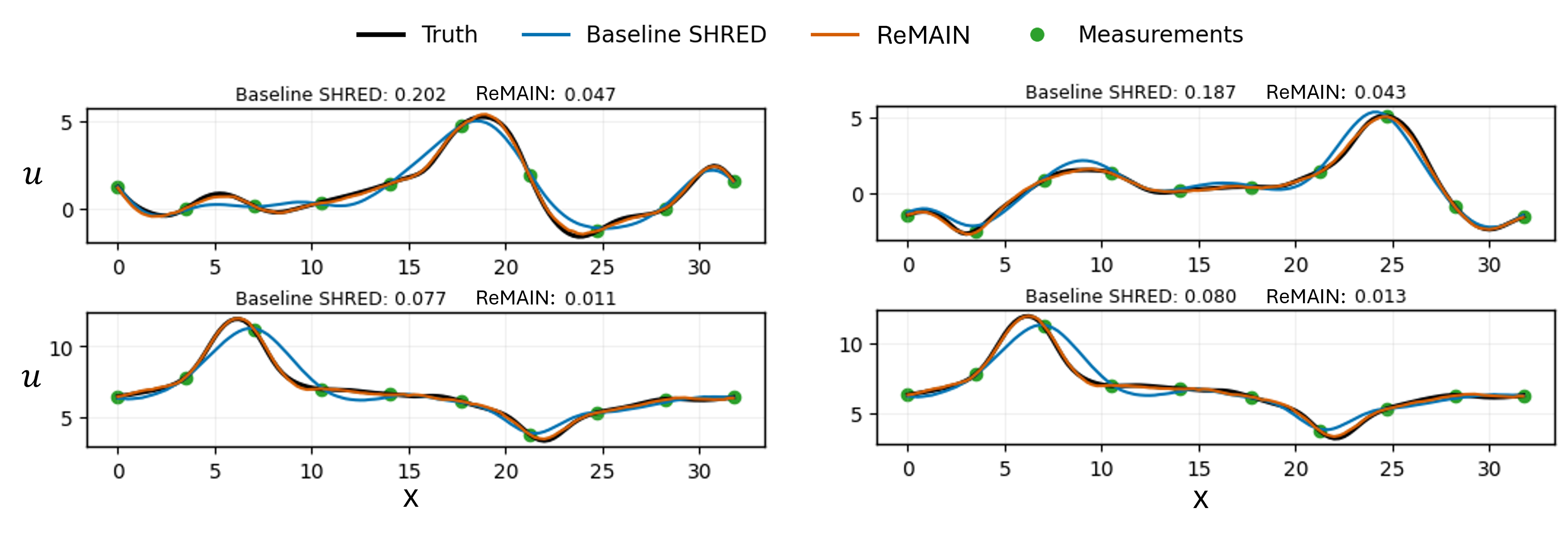}
\caption{Comparison of SHRED and ReMAIN reconstructions for representative held-out test snapshots of the driven Sine-Gordon system. Values above each subplot indicate the relative \(L_{2}\) reconstruction errors of SHRED and ReMAIN for the displayed snapshot.}
\label{fig:8}
\end{figure}

\begin{figure}[htbp]
\centering
\includegraphics[width=1\linewidth,height=0.78\textheight,keepaspectratio]{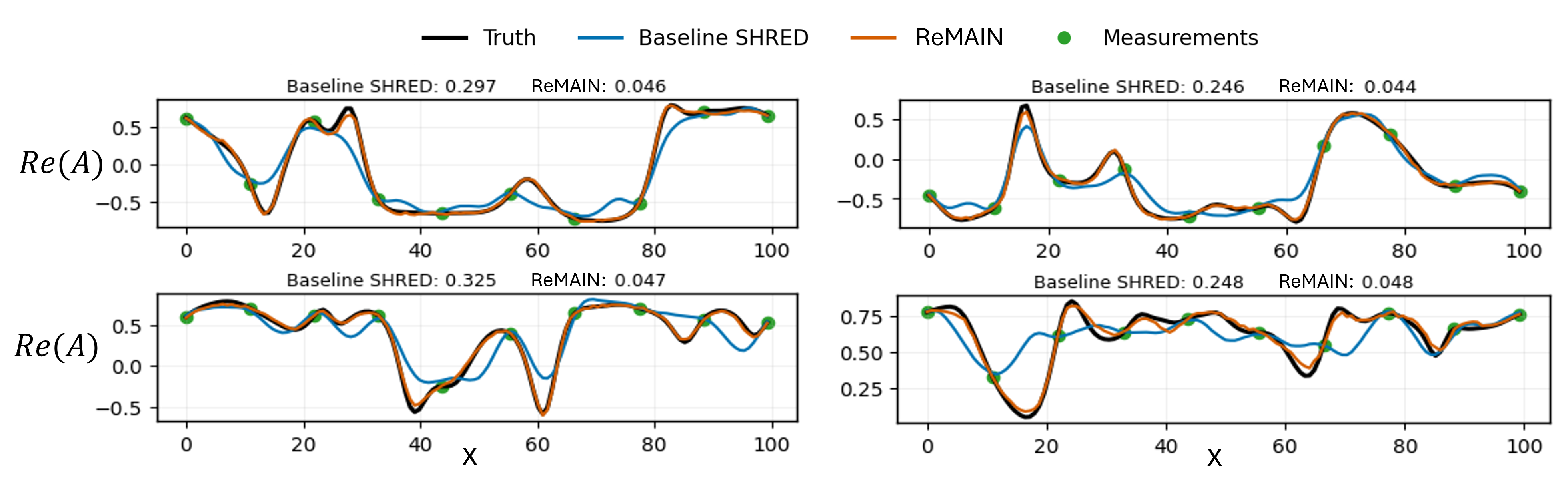}
\caption{Comparison of SHRED and ReMAIN reconstructions for representative held-out test snapshots of the driven Complex Ginzburg--Landau system. Values above each subplot indicate the relative \(L_{2}\) reconstruction errors of SHRED and ReMAIN for the displayed snapshot.}
\label{fig:9}
\end{figure}

Representative test snapshots in Figures~\ref{fig:4}--\ref{fig:9} were selected to highlight cases where the difference between the two reconstruction approaches is visually apparent, clearly demonstrating the improved reconstruction obtained with ReMAIN. The improvement is particularly apparent for solutions containing sharp transition fronts, localized extrema, or more complex spatial variation. In such regions, SHRED commonly smooths sharp features or introduces errors in their position and amplitude, whereas ReMAIN more faithfully reproduces the location and shape of the underlying structures. This distinction is especially pronounced for the sharp transitions in the compressible Navier--Stokes and FitzHugh--Nagumo cases (Figures~\ref{fig:6},\ref{fig:7}), where ReMAIN preserves the front position and steepness more faithfully. For the KdV--KS, sine--Gordon, and CGL cases (Figures~\ref{fig:4}, \ref{fig:8}, and \ref{fig:9}), ReMAIN also better captures localized peaks, troughs, and finer spatial variations. The forced Burgers results (Figure~\ref{fig:5}) show the same tendency, although the smoother spatial profiles make the difference between the two models less pronounced. Overall, these examples indicate that the principal benefit of ReMAIN lies in recovering spatial structure that is degraded or displaced by the shallow fully connected decoder in the SHRED architecture.

\begin{figure}[p]
\centering
\includegraphics[width=1\linewidth,height=0.85\textheight,keepaspectratio]{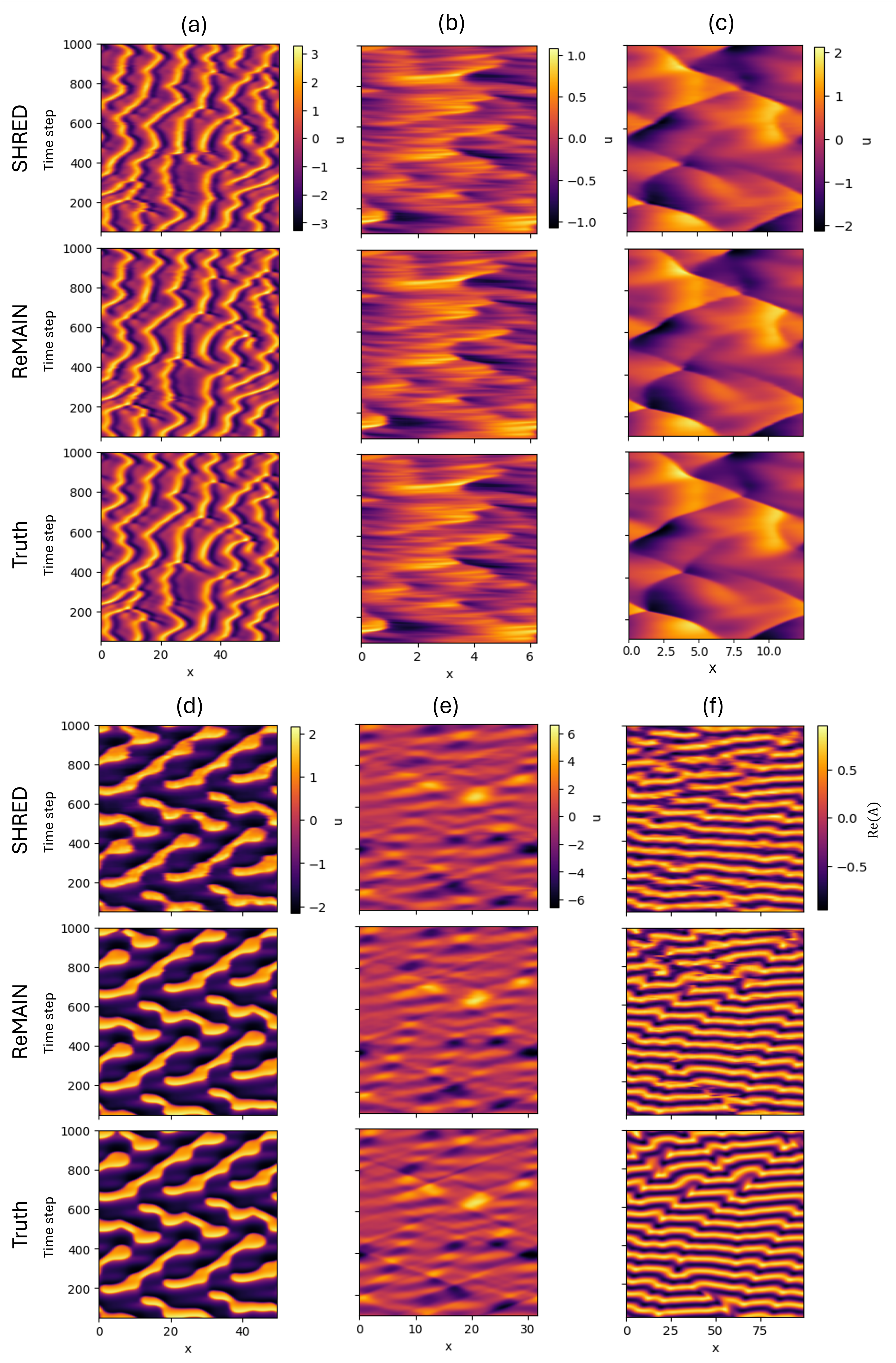}
\caption{Comparison of SHRED and ReMAIN reconstructions of spatiotemporal evolution for a sample test trajectory across the six PDE benchmarks; (a) KdV--KS, (b) Forced Burgers, (c) Compressible Navier-Stokes, (d) FitzHugh-Nagumo, (e) Sine-Gordon, and (f) Ginzburg--Landau equation.}
\label{fig:10}
\end{figure}

\begin{figure}[p]
\centering
\includegraphics[width=.86\linewidth,height=0.78\textheight,keepaspectratio]{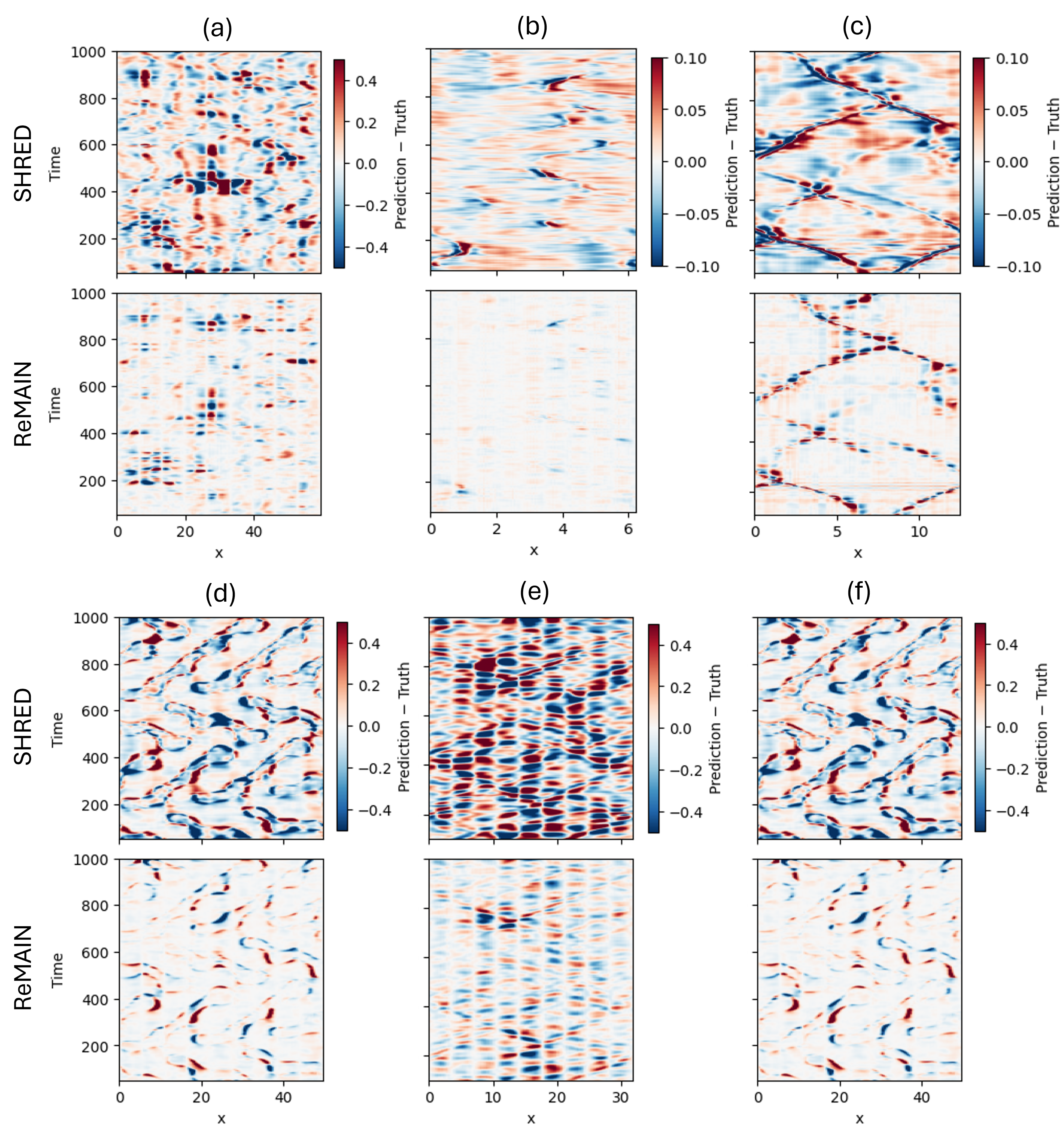}
\caption{Difference between SHRED and ReMAIN reconstructions and the ground truth for the sample test trajectories shown in Fig.~\ref{fig:10} across the six PDE benchmarks; (a) KdV--KS, (b) Forced Burgers, (c) Compressible Navier--Stokes, (d) FitzHugh--Nagumo, (e) Sine--Gordon, and (f) Ginzburg--Landau equation.}
\label{fig:11}
\end{figure}

\FloatBarrier
\begin{samepage}
The sample spatiotemporal trajectories in Figure~\ref{fig:10} show that both models reproduce the dominant temporal evolution of the PDE systems. Figure~\ref{fig:11} makes the distinction clearer by revealing where the reconstruction error is concentrated: for SHRED, the residuals form stronger, wider bands around sharp transitions and the boundaries of evolving structures, indicating localized errors in feature position and shape. These concentrated residuals are substantially weaker for ReMAIN, showing that it more effectively suppresses errors where the solution varies most rapidly in space and time.
\end{samepage}

\FloatBarrier
\section{Application of parametric ReMAIN to a plasma problem}

The parametric demonstration considers a collisionless \(E \times B\) plasma discharge representative of the radial--azimuthal cross-section of a Hall thruster. The configuration is two-dimensional in the radial \(x\) and azimuthal \(z\) directions, with an externally imposed axial electric field \(E\) and radial magnetic field \(B\). The plasma dynamics are governed by coupled instabilities, principally the electron cyclotron drift instability (ECDI) and modified two-stream instability (MTSI), whose relative importance and spatial structure vary substantially with the applied electric and magnetic field strengths \cite{r06,r07}. The problem therefore provides a demanding parameter-dependent reconstruction test in which changes in the operating parameter alter the underlying spatiotemporal plasma dynamics rather than simply rescaling the same solution.

Unlike the canonical PDE benchmarks of Section 3, the plasma dataset is generated using a first-principles Particle-in-Cell (PIC) simulation \cite{dawson1983,taccogna2023}, rather than by direct integration of prescribed macroscopic PDEs. In PIC, the plasma is represented by an ensemble of Lagrangian computational particles that evolve continuously through space, while macroscopic plasma quantities are obtained by depositing particle velocity moments onto a computational grid. The resulting fields therefore contain finite-particle statistical fluctuations superimposed on the underlying physical multiscale dynamics. The plasma case thus combines strongly nonlinear, regime-dependent dynamics with the statistical variability and noise inherent to particle-based kinetic data.

The data were generated using the Reduced-Order quasi-two-dimensional PIC method \cite{r23}, under its 50-region approximation \cite{r24}. The method is extensively verified for \(E \times B\) plasmas \cite{r23,r24,r25}. The simulation setup follows the established radial--azimuthal Hall-thruster benchmark configuration \cite{r26}. The physical domain extends \(1.28cm\ \) in both radial and azimuthal directions. The baseline axial electric field is \(E_{0} = 10kV/m\), and the applied radial magnetic field is fixed at \(B = 20mT\). Electrons and xenon ions are initially sampled from Maxwellian distributions with temperatures of \(10\) and \(0.5eV\), respectively, at an initial plasma particle number density of \(1.5 \times 10^{16}\ m^{- 3}\). The simulations are collisionless and are advanced for \(30\mu s\) with a numerical time step of \(1.5 \times \ 10^{- 11}s\).

Simulations were executed over a range of normalized axial electric-field strengths as operating conditions. Simulation data with \(\frac{E}{E_{0}} = \left\{ 1.0,\ 1.5,\ 2.0,\ 2.5,\ 3.0,\ 3.5,\ 4.0,\ 4.5,\ 5.0 \right\}\) \cite{r06} were used for training. Generalization in operating-parameter space was evaluated using entirely held-out test simulation data with \(\frac{E}{E_{0}} = \left\{ 1.25,\ 1.75\ ,2.25,\ 2.75,\ 3.25,\ 3.75,\ 4.25,\ 4.75 \right\}\). For the present reconstruction problem, the snapshots of the axial electron-current-density field, \(J_{ey}\), were used. Outputs represent the average of \(1000\) consecutive simulation snapshots of the macroscopic plasma properties, yielding \(2000\) stored snapshots for each simulation. The first \(350\) outputs, corresponding to the initial \(5\mu s\) transient, were discarded, leaving \(1650\) snapshots for each operating-parameter regime. Each was represented on a \(50 \times 256\) radial--azimuthal grid, giving 12,800 spatial values per snapshot.

\begin{figure}[htbp]
\centering
\includegraphics[width=.31\linewidth,height=0.78\textheight,keepaspectratio]{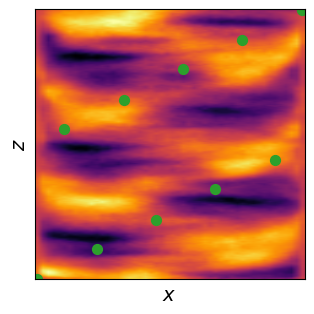}
\caption{Spatial distribution of the ten fixed sensor locations over the radial--azimuthal \(J_{ey}\) field; green markers indicate the sensor positions.}
\label{fig:12}
\end{figure}

Ten fixed, equidistant spatial locations over the flattened radial--azimuthal grid were used to represent sparse sensors measuring \(J_{ey}\), leading to the spatial arrangement shown in Figure~\ref{fig:12}. Each model input comprised a 50-snapshot history of the measurements from these sensor locations.

Measurement and target fields were normalized using a min--max scaler fitted jointly across the full training dataset, encompassing all training parameter regimes. The operating parameter \(E/E_{0}\) was standardized using the mean and standard deviation of the training parameter values. The same normalization and standardization parameters obtained from the training data were used to transform the held-out test dataset.

\subsection{Model settings and training procedure}

For the plasma case, parameter-conditioned ReMAIN used a two-layer LSTM with 64 hidden units and a two-layer fully connected parameter encoder with 16 units per layer and GELU activations, producing a 16-dimensional parameter embedding. Together, the temporal and parameter encoders create an 80-dimensional joint history--parameter representation. The two-dimensional FiLM U-Net used a base width of \(32\), corresponding to 32--64--128--256 channels across the encoder and bottleneck. Radial and azimuthal coordinates, normalized to \(\lbrack - 1,1\rbrack,\) were used as the two spatial coordinate channels.

Training used batches of 16 for a maximum of \(300\) epochs, with model performance evaluated every \(10\) epochs. In addition to the Smooth-\(L_{1}\) reconstruction loss used for the PDE cases, an auxiliary multiscale loss was introduced to improve preservation of the spatial structures characteristic of the plasma dynamics. This term measured the relative squared error of signed azimuthal field increments at separations of \(1,\ 2,\ 4\), and \(8\) grid points after inverse scaling to physical units, with weight \(0.02\). It was inactive for the first 50 epochs and increased linearly to its full weight over the following \(50\) epochs. All remaining optimizer, learning-rate scheduling, and gradient-clipping settings remained identical to those described for the PDE benchmarks in Section 3.1.

\subsection{Results}

Reconstruction performance is quantified independently for each held-out operating parameter \(E\). For a given test parameter, all retained test snapshots were evaluated individually, producing a distribution of snapshot-level errors. The bar plots presented later in this section report the mean of these errors, with error bars denoting the population standard deviation across snapshots.

Three complementary error metrics are used. The full-field relative \(L_{2}\) error, used for PDE benchmark problems, provides a direct pointwise measure of reconstruction accuracy, but it can over-penalize small spatial displacements, phase offsets, or local deformations of coherent multiscale structures: two fields may be very similar in their overall morphology and scale content while exhibiting a comparatively large pointwise error because corresponding structures are not perfectly aligned. To distinguish such pointwise discrepancies from errors in the underlying multiscale structure, we therefore additionally quantify the azimuthal spectral error and the second-order structure-function error, which assess agreement in the distribution of spatial scales and in real-space fluctuation structure, respectively.

\textbf{Azimuthal spectral error.} This metric measures how accurately the reconstruction reproduces the relative distribution of fluctuation power across azimuthal spatial wavenumbers, by calculating the azimuthal power spectrum for each snapshot. At each radial position \(i\), the azimuthal mean was removed, and a one-dimensional Fourier transform was applied along the azimuthal grid direction \(j\) yielding the normalized one-sided Fourier coefficient \({\widehat{u}}_{k}(i)\) for azimuthal wavenumber \(k\). The corresponding modal power was then averaged over all radial positions to obtain the azimuthal power spectrum,

\begin{equation}
\label{eq:30}
P(k) = \frac{1}{N_{i}}\sum_{i = 1}^{N_{i}}\left| {\widehat{u}}_{k}(i) \right|^{2},\ \ \ \ \widetilde{P}(k) = \frac{P(k)}{\sum_{k}^{}P(k)},
\end{equation}

The error was then evaluated in logarithmic space as

\begin{equation}
\label{eq:31}
\epsilon_{spec} = \left\lbrack \frac{1}{N_{k}}\sum_{k}^{}\left( \log_{10}\left\lbrack {\widetilde{P}}_{pred}(k) \right\rbrack - \log_{10}\left\lbrack {\widetilde{P}}_{truth}(k) \right\rbrack \right)^{2} \right\rbrack^{1/2},
\end{equation}

The error was evaluated in logarithmic space to prevent the error metric from being dominated by the highest-energy spectral modes and gives greater sensitivity to differences across the full resolved spectral range.

\textbf{Second-order structure function error.} This metric measures how accurately the reconstruction reproduces the scale-dependent spatial organization of fluctuations in physical space. For each snapshot, the second-order structure function is defined as

\begin{equation}
\label{eq:32}
S_{2}(r) = \left\langle \left\lbrack u(x + r) - u(x) \right\rbrack^{2} \right\rangle_{x}\ 
\end{equation}

where \(r\) is the spatial separation vector and \(\left\langle . \right\rangle_{x}\) denotes averaging over all spatial locations. In the discrete two-dimensional grid, the structure function was evaluated for separations \(r = \{ 1,2,4,8,16\}\) grid spacings along each coordinate direction and averaged over the two directions,

\begin{equation}
\label{eq:33}
S_{2}(r) = \frac{1}{2}\left\lbrack \left\langle (u_{i + r,j} - u_{i,j})^{2} \right\rangle_{i,j} + \left\langle (u_{i,j + r} - u_{i,j})^{2} \right\rangle_{i,j} \right\rbrack.
\end{equation}

Periodic shifts were used so that points displaced beyond a domain boundary re-enter from the opposite side, allowing all grid points to contribute equally. To remove dependence on the overall fluctuation amplitude, the structure function was normalized by twice the spatial variance of the same snapshot,

\begin{equation}
\label{eq:34}
{\widetilde{S}}_{2}(r) = \frac{S_{2}(r)}{2\, \operatorname{Var}(u)}.
\end{equation}

The error was then calculated over the set of retained lags as

\begin{equation}
\label{eq:35}
\epsilon_{S_{2}} = \frac{\left\| {\widetilde{S}}_{2,pred} - {\widetilde{S}}_{2,truth} \right\|_{2}}{\left\| {\widetilde{S}}_{2,truth} \right\|_{2}}.
\end{equation}

\begin{figure}[htbp]
\centering
\includegraphics[width=.85\linewidth,height=0.78\textheight,keepaspectratio]{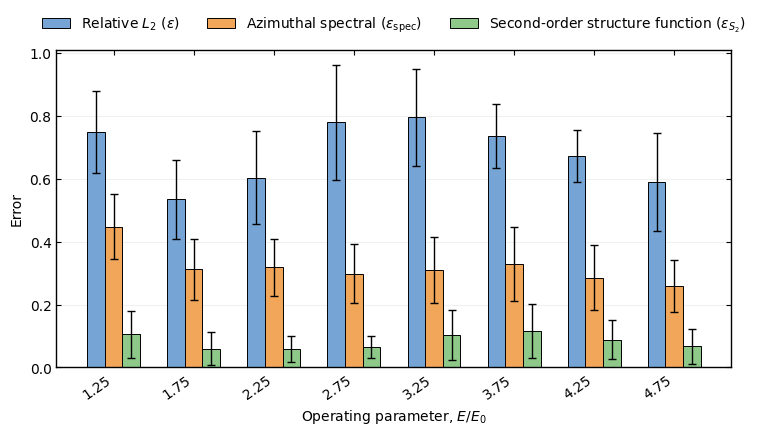}
\caption{Mean relative \(L_{2}\), azimuthal spectral, and second-order structure-function errors of parametric ReMAIN reconstructions for the plasma test case. The error metrics are averaged over all snapshots for each test operating parameter, and the error bars show the corresponding population standard deviation across snapshots.}
\label{fig:13}
\end{figure}

Using only the sparse measurements at the ten fixed locations shown in Figure~\ref{fig:12}, parameter-conditioned ReMAIN reconstructs the plasma dynamics across all held-out \(E/E_{0}\) conditions. The three metrics in Figure~\ref{fig:13} provide complementary views of this performance. While the pointwise relative \(L_{2}\) error remains large for the reason stated earlier, the substantially lower spectral and second-order structure-function errors show that the model preserves the distribution and spatial organization of fluctuation scales even when individual coherent structures are not perfectly aligned with the ground truth. Importantly, no systematic degradation is observed across the tested operating-parameter range, indicating that reconstruction quality is retained as the underlying plasma dynamics vary significantly across the parameter regime.

\begin{figure}[htbp]
\centering
\includegraphics[width=1\linewidth,height=0.78\textheight,keepaspectratio]{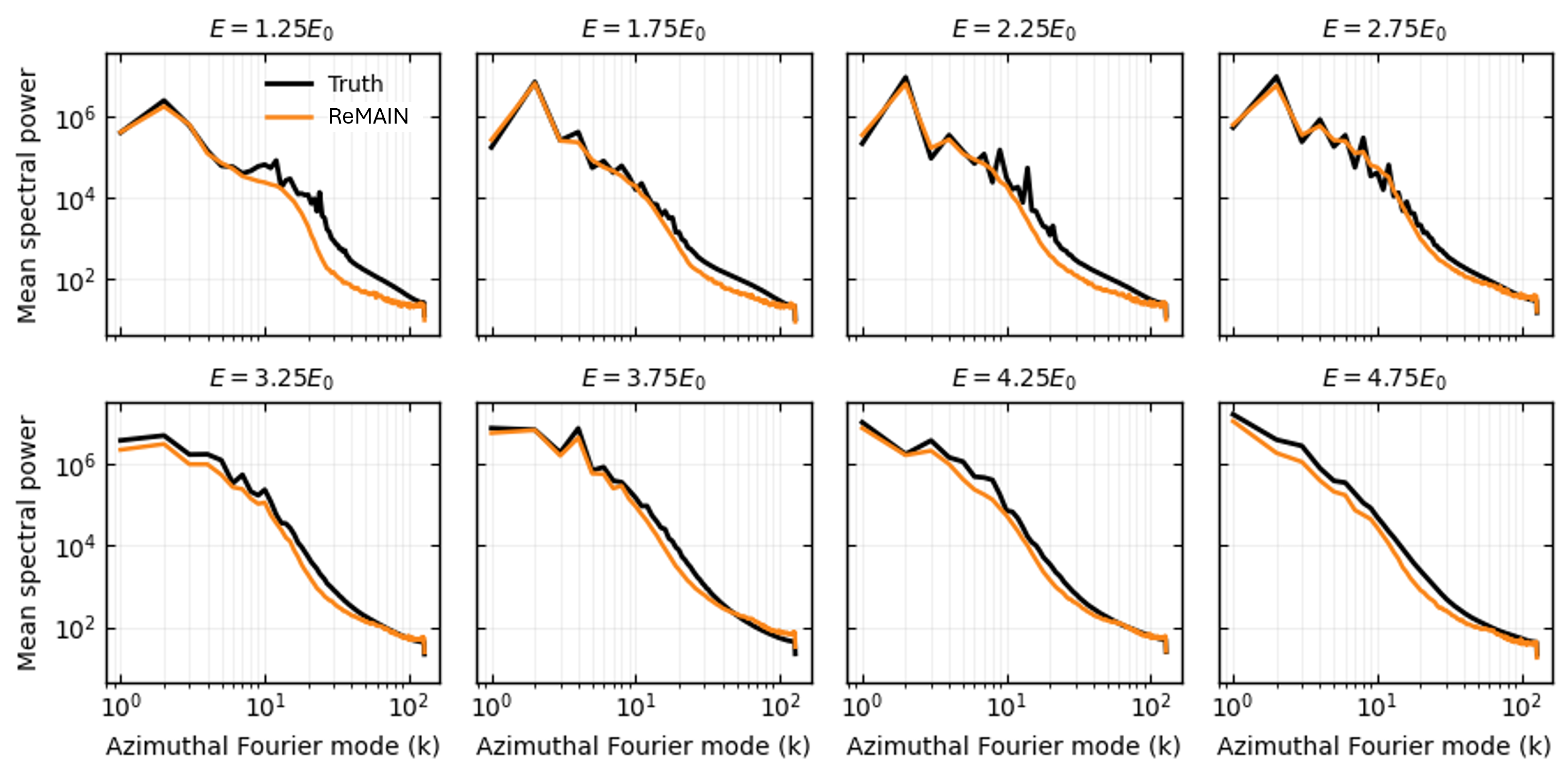}
\caption{Mean azimuthal power spectra of truth and parametric ReMAIN across test operating parameters. Note both axes are shown on logarithmic scales.}
\label{fig:14}
\end{figure}

This interpretation is further supported by the mean azimuthal power spectra in Figure~\ref{fig:14}, which characterize the spatial-wavenumber content of the reconstructed fields. ReMAIN closely reproduces the parameter-dependent spectral distributions across the tested operating conditions, suggesting that it captures both the characteristic spatial scales of the plasma fluctuations and how their relative contributions vary with the operating parameter.

\begin{figure}[p]
\centering
\includegraphics[width=1\linewidth,height=0.85\textheight,keepaspectratio]{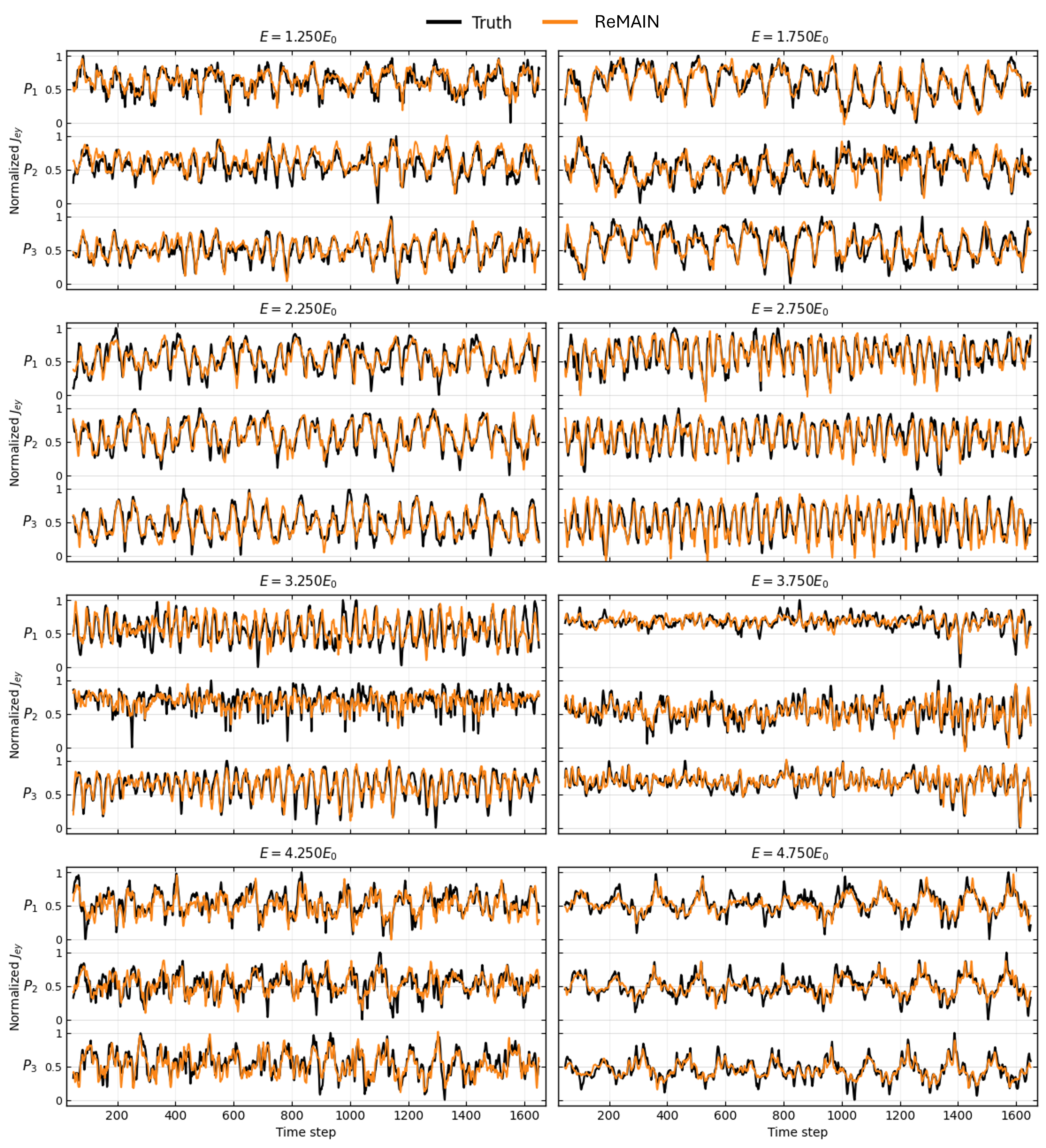}
\caption{Comparison of ground truth and parametric ReMAIN temporal evolution at three randomly selected spatial locations for each test operating parameter; each signal is independently normalized using the corresponding ground-truth range.}
\label{fig:15}
\end{figure}

\begin{figure}[p]
\centering
\includegraphics[width=1\linewidth,height=0.85\textheight,keepaspectratio]{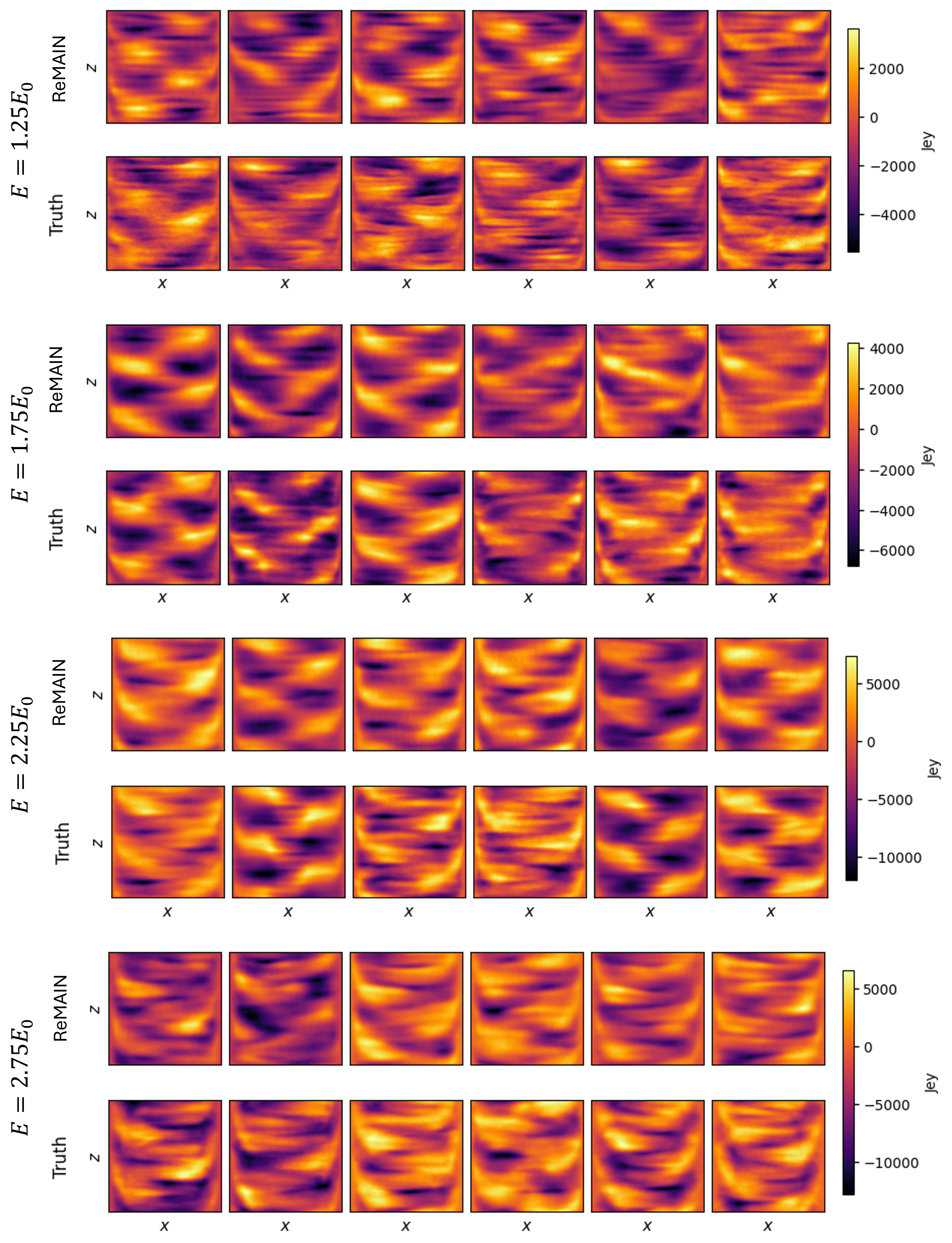}
\caption{Comparison of parametric ReMAIN and ground-truth snapshots for sample test snapshots at \(\frac{E}{E_{0}} = 1.25,\ 1.75,\ 2.25,\) and \(\ 2.75\).}
\label{fig:16}
\end{figure}

\begin{figure}[p]
\centering
\includegraphics[width=1\linewidth,height=0.85\textheight,keepaspectratio]{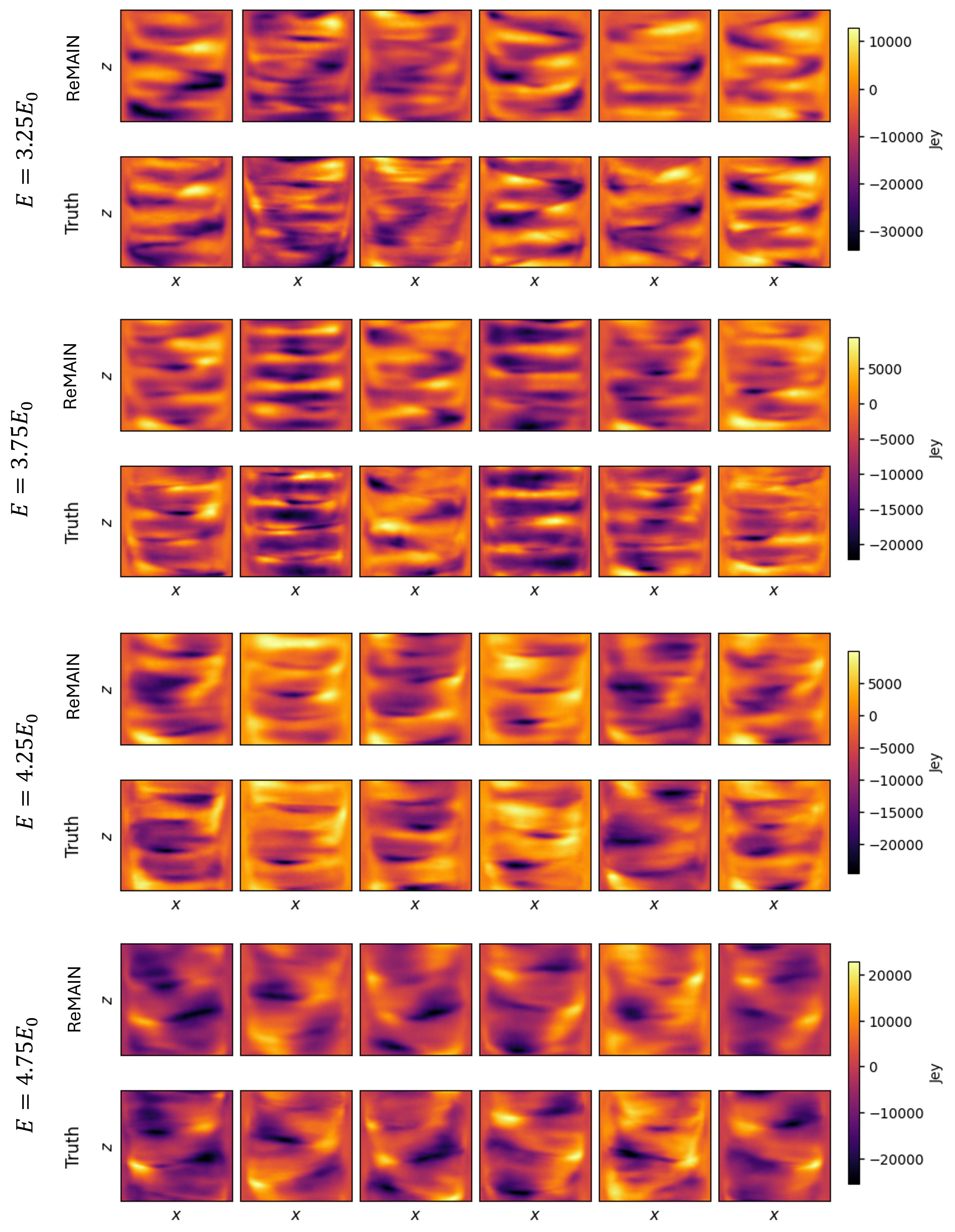}
\caption{Comparison of parametric ReMAIN and ground-truth snapshots for sample test snapshots at \(\frac{E}{E_{0}} = 3.25,\ 3.75,\ 4.25,\) and \(\ 4.75\).}
\label{fig:17}
\end{figure}

The temporal signals in Figure~\ref{fig:15} additionally demonstrate that the reconstructed fields retain the local temporal evolution over the full test intervals at random spatial locations. Finally, the sample two-dimensional snapshots in Figure~\ref{fig:16} and Figure~\ref{fig:17} likewise illustrate that ReMAIN captures the parameter-dependent morphology, orientation, and spatial scale of the \(J_{ey}\) structures. The dominant coherent structures are recovered, with the remaining discrepancies appearing mainly as local amplitude differences and small spatial or phase offsets. These residuals should be interpreted in the context of finite-history state observability. Reconstruction of a \(12,800\)-dimensional field from a finite history of measurements at only ten spatial locations is strongly underdetermined unless the observed history uniquely constrains all unmeasured degrees of freedom through the system dynamics. Although temporal evolution couples observed and unobserved degrees of freedom and thereby provides substantially more information than instantaneous measurements alone, information from ten sensor histories does not necessarily define a unique mapping to the 12,800-dimensional plasma state. Spatially localized or rapidly varying degrees of freedom may leave weak signatures at the sensors, so multiple dynamically admissible states can remain nearly indistinguishable under the available observations while differing locally in structure position, phase, or amplitude. A deterministic reconstruction such as ReMAIN necessarily selects a single realization from this remaining conditional uncertainty; exact pointwise agreement would therefore require the measurement history to constrain essentially all relevant state-space directions \cite{r13,r18}. Consequently, small positional or phase differences can generate appreciable pointwise error, particularly across sharp gradients, while leaving the characteristic multiscale organization largely preserved. The contrast between the pointwise, spectral, and structure-function errors in Figure~\ref{fig:13} is consistent with this interpretation. Overall, the results demonstrate that parameter-conditioned ReMAIN generalizes to unseen operating conditions within the range spanned by the training parameters while retaining the characteristic temporal and multiscale spatial organization of the plasma dynamics.

\FloatBarrier
\section{Conclusion}

This work introduced ReMAIN, a recurrent multiscale framework for reconstructing high-dimensional dynamical states from sparse measurement histories. ReMAIN retains the central SHRED principle of extracting information from temporal sensor histories, while replacing its shallow fully connected decoder with a FiLM-conditioned multiresolution U-Net. The recurrent representation serves both as a dense prior for the reconstructed state and as scale-specific conditioning that modulates feature processing throughout the U-Net hierarchy. In the parametric extension, the operating condition is embedded jointly with the measurement history, allowing the reconstruction to adapt as the underlying dynamics change across regimes.

Across the six canonical nonlinear PDE problems spanning substantially different spatiotemporal behaviours, ReMAIN consistently outperformed SHRED on test trajectories, reducing the mean relative \(L_{2}\) reconstruction error by approximately 25--80\%, depending on the system. The improvement was most pronounced for sharp fronts, localized extrema, and finer-scale spatial variations, which ReMAIN reconstructed with greater fidelity in their position, shape, and amplitude. The parameter-conditioned extension was further demonstrated for a substantially higher-dimensional \(E \times B\) plasma problem in which changes in the applied electric field modify the underlying instability dynamics. ReMAIN successfully reconstructed the plasma state while preserving its characteristic dynamics, spectral content, and scale-dependent spatial organization, including their variation across operating conditions withheld from training.

ReMAIN's capability is particularly relevant to \(E \times B\) plasma systems, where experimentally accessible diagnostics often provide limited or localized measurements of an evolving discharge, while applications such as monitoring and control require richer knowledge of the plasma state. ReMAIN offers a route for augmenting existing diagnostics with estimates of otherwise unobserved spatial structure, inferring quantities and regions that are inaccessible or costly to measure directly, and providing rapid state estimates for online monitoring, feedback control, and data-driven digital-twin workflows. In this role, high-fidelity simulations would remain essential for model training and validation, while the trained model could provide computationally inexpensive state estimates during operation. Practical deployment will require the diagnostic inputs, reconstructed quantities, and training regimes to be selected jointly so that the measurements are both experimentally feasible and dynamically informative of the target state. Uncertainty quantification and out-of-distribution detection will also be important for identifying when reconstructions become unreliable during transients, regime changes, faults, or diagnostic degradation. Beyond state reconstruction, the framework could support diagnostic design by quantifying how reconstruction accuracy varies with sensor number and placement, helping balance information content against accessibility, cost, survivability, and intrusiveness. Together, the results establish ReMAIN as a promising step toward multiscale, regime-aware full-state inference from highly limited measurements in complex plasma systems.

\section*{Data availability}
The ReMAIN code will be made publicly available upon publication.

\bibliographystyle{unsrt}
\bibliography{references}
\end{document}